\documentclass[ 12pt]{article}
\usepackage{authblk}
\usepackage{ae}
\usepackage[T1]{fontenc}
\usepackage[ansinew]{inputenc}
\usepackage{amsmath}
\usepackage{amssymb}
\usepackage{color}
\usepackage{graphicx}
\usepackage{longtable}
\usepackage{latexsym}
\usepackage{stmaryrd}
\usepackage{amsfonts}
\usepackage{amsbsy}
\usepackage{mathrsfs}
\usepackage{psfrag}
\usepackage{enumerate}
\usepackage{MnSymbol}
\usepackage{vmargin}
\setmarginsrb{3cm}{2cm}{3cm}{2cm}{1cm}{1cm}{1cm}{1cm}

\usepackage{cite}

\usepackage{epstopdf}
\usepackage{amsmath,amssymb,calc,amsfonts}
\usepackage{latexsym}

\usepackage{tikz}
\usetikzlibrary{matrix}

\usepackage{color}

\usepackage{graphicx,calc,epsfig}

\newcommand{\C}{\mathbb{C}}
\newcommand{\R}{\mathbb{R}}

\newcommand{\n}{\nonumber}
\newcommand{\be}{\nopagebreak[3]\begin{equation}}
\newcommand{\ee}{\end{equation}}
\newcommand{\bee}{\nopagebreak[3]\begin{equation*}}
\newcommand{\eee}{\end{equation*}}
\newcommand{\ba}{\nopagebreak[3]\begin{eqnarray}}
\newcommand{\ea}{\end{eqnarray}}
\newcommand{\baa}{\nopagebreak[3]\begin{eqnarray*}}
\newcommand{\eaa}{\end{eqnarray*}}
\newcommand{\la}{\label}
\DeclareFontFamily{U}{rsfs}{}         
\DeclareFontShape{U}{rsfs}{m}{n}{<5> rsfs5 <6><7> rsfs7          %
  <8><9><10><10.95><12><14.4><17.28><20.74><24.88> rsfs10}{}     %
\DeclareMathAlphabet{\mathfs}{U}{rsfs}{m}{n}                     %
\newcommand{\mfs}[1]{\mathfs {#1}}                               %

\newcommand{\va}{\scriptscriptstyle}

\newcommand{\sH}{{\mfs H}}

\newcommand{\sM}{{\mfs M}}

\newcommand{\Hk}{{\sH}_{kin}}

\newcommand{\su}{\mathfrak{su}}
\newcommand{\so}{\mathfrak{so}}

\newcommand{\tr}{\mathrm{tr}}

\begin{document}

\title{Symmetries of quantum space-time in 3 dimensions}
\author[1]{Francesco Cianfrani\thanks{francesco.cianfrani@ift.uni.wroc.pl}}
\author[2]{Jerzy Kowalski-Glikman\thanks{jerzy.kowalski-glikman@ift.uni.wroc.pl}}
\author[3]{Daniele Pranzetti\thanks{dpranzetti@sissa.it}}
\author[4]{Giacomo Rosati\thanks{giacomo.rosati@ift.uni.wroc.pl}}
\affil[1,2,4]{Institute for Theoretical Physics, University of Wroc\l{}aw, Pl.\ Maksa Borna
9, Pl--50-204 Wroc\l{}aw, Poland}
\affil[3]{Scuola Internazionale Superiore di Studi Avanzati (SISSA), via Bonomea 265, 34136 Trieste, Italy}
\sloppy
\maketitle
\pagestyle{plain}

\begin{abstract}

By applying loop quantum gravity techniques to 3D gravity with a positive cosmological constant $\Lambda$, we show how the local gauge symmetry of the theory, encoded in the constraint algebra, acquires the quantum group structure of $so_q(4)$, with $ q = \exp{(i\hbar \sqrt{\Lambda}/2\kappa)}$. By means of an Inonu-Wigner contraction of the quantum group bi-algebra, keeping $\kappa$ finite, we obtain the kappa-Poincar\'e algebra of the flat quantum space-time symmetries.
\end{abstract}

\sloppy
\maketitle

\pagestyle{plain}
\newpage{}

\section{Introduction}\la{Introduction}

What are the symmetries of flat quantum space-time? Given the fundamental role the Poincar\'e group -- the group of symmetries of the flat Minkowski space-time -- plays in quantum field theory, answering this question is most likely a prerequisite for understanding physics at the Planck scale and beyond. It is widely believed that quantum gravity changes dramatically the space-time structure at small distances, allowing for fluctuations of space-time itself. One could expect therefore that, as a result of the presence of these quantum fluctuations, the symmetries of such quantum space-time, even in the flat limit, where spacetime curvature is negligible, in some sense should differ from the classical Poincar\'e symmetries. Over many years more and more circumstantial evidences has accumulated indicating that the symmetries of flat quantum space-time should be described by some kind of quantum deformation of the Poincar\'e group, and that the infinitesimal generators of the deformed group are to be described by a Hopf algebra, which is a deformation of the Poicar\'e algebra.

One of the possible Hopf-deformed Poincar\'e algebras is the so-called $\kappa$-Poincar\'e algebra, constructed twenty five years ago by Lukierski, Nowicki, Ruegg and Tolstoy in \cite{Lukierski:1991pn}-\cite{Lukierski:1992dt} and brought into its final form in \cite{Majid:1994cy}. It turns out that the construction of this algebra is highly nontrivial due to the fact that the classical Poincar\'e algebra, which is a semidirect product of Lorentz algebra and a commuting algebra of translations, is not simple and the standard Drinfeld-Jimbo methods of constructing the deformed algebras does not work here. Instead the authors of \cite{Lukierski:1991pn}-\cite{Lukierski:1992dt} took as their starting point the deformed Anti de Sitter algebra (in four space-time dimensions) $\so_q(3,2)$ and then performed its contraction, by going to the limit of vanishing cosmological constant. As we will see below even taking such a contraction limit is not completely straightforward.

The construction of $\kappa$-Poincar\'e algebra sparked  a lot of interest and many of its properties have been investigated (in particular in the context of quantum gravity phenomenology, see \cite{AmelinoCamelia:2008qg} and references therein), but the most fundamental question as to whether $\kappa$-Poincar\'e indeed has anything to do with quantum space-time symmetries has never been answered in a satisfactory way. In this paper we would like to do so in the context of a toy model of quantum gravity in 3 space-time dimensions.

Since the steps of our argument, which will be presented in the following sections, are pretty technical, we will present here an extensive overview of what we are going to do. The reader is referred to the rest of this paper for more detailed technical discussion.

The starting point of the next section is the algebra of constraints of classical general relativity in 3 dimensions. It consists of two generators of spatial diffeomorphism ${\cal D}_i(x)$ and Hamiltonian constraint ${\cal H}(x)$. One can integrate these constraints on a spacelike surface with some smearing functions,
 \begin{equation}\label{j1}
{\cal D}[f] = \int d^2x\, f^a(x) {\cal D}_a(x) \,,\quad   {\cal H}[g] = \int d^2x\, g(x) {\cal H}(x)
\end{equation}
obtaining in this way an equivalent description of the constraints. Their Poisson algebra  is not a Lie algebra however, because on the right hand side of the brackets we have to deal with a {\em metric dependent} structure function, instead of  structure constants.
However, if we choose the smearing functions $N^i(x)$ and $M(x)$ to be linear in $x$, and assume that the space-time metric is the flat Minkowski one, the algebra of  constraints (\ref{j1}) becomes the Poincar\'e algebra. This is easy to understand, because in this case the smeared constraint become a natural generalization of the Poincar\'e generators of translations, boosts and rotation. If, instead, we make the smearing functions equal to the components of the Killing vectors of (Anti) de Sitter space, and substitute the corresponding (Anti) de Sitter space metric on the right hand side, we obtain an algebra isomorphic to $\so(3,1)$ (or $\so(2,2)$) in the case of the positive (or negative) cosmological constant.

In this paper, for technical reasons, we will work in Euclidean spaces, but the results described above hold here as well, and the algebra of smeared constraints with appropriate smearing functions becomes isomorphic to $\mathfrak{iso}(3)$ for $\lambda=0$, $\so(4)$ for $\Lambda>0$ and $\so(3,1)$ (with compact boosts) for $\Lambda <0$.

The above concerned the metric formulation of $3D$ gravity. In the Chern-Simons formulation of this theory instead of the diffeomorphism and Hamiltonian constraints we have to deal with six constraints corresponding to a gauge symmetry group, which is\footnote{From now on we restrict ourselves to the case of the Euclidean, positive (or zero) cosmological constant that we are going to analyze in this paper.} $ISO(3)$ in the case of vanishing cosmological constant and $SO(4)$ in the case of  positive $\Lambda$. The relation between these constraints and the ones of the metric formulation is highly nontrivial in general, but, as explained in Sections \ 2 and 3, these two sets of constraints become essentially identical, if we assume that the metric of space time is maximally symmetric.

We conclude therefore that {\em the algebra of gauge constraints is the algebra of space-time symmetries in the case of (Euclidean) flat or de Sitter space-times}. The aim of this paper is to employ this identification on the quantum level, in order to find out what are the symmetries of {\em quantum} de Sitter and flat Euclidean spaces.

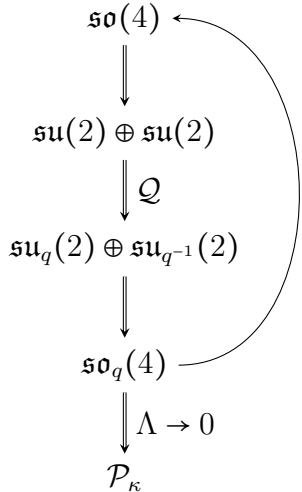
\begin{figure}
\begin{center}
\begin{tikzpicture}
  \matrix (m) [matrix of math nodes,row sep=2em,column sep=4em,minimum width=2em]
  {
     \so(4) \\
     \su(2) \oplus \su(2)  \\
     \su_{q}(2) \oplus \su_{q^{-1}}(2) \\
     \so_q(4) \\
     {\cal P}_\kappa
    \\};
  \path[-stealth]
    (m-1-1) edge [double]  (m-2-1)
    (m-2-1) edge [double] node [right] {${\cal Q}$} (m-3-1)
    (m-3-1) edge [double]  (m-4-1)
    (m-4-1) edge [double] node [right] {$ \Lambda \rightarrow 0$} (m-5-1)
    (m-4-1) edge [out=0,in=0,looseness = 1.2] (m-1-1)
            ;
\end{tikzpicture}
\caption{The logical steps of the paper}
\end{center}
\end{figure}

The logic of this construction is depicted in Fig.\ 1. The starting point  is  the set of gauge constraints of Eucledean $3D$ gravity with positive cosmological constant. We  include the cosmological constant from the very start because it serves as a necessary infrared regulator for the quantization procedure, rendering the gauge group manifold compact. In Sects.\ 2 and 3 we also make use of the decomposition of the  algebra $\so(4)$ into the direct sum of two $\su(2)$ algebras, decomposing the gauge generators appropriately. In this way we obtain the two commuting algebras of gauge generators, which we consider separately, for a while.

In Sect.\ 4 we take the first step towards the central result of our paper, the route from the classical algebra of $\so(4)$ to its quantum counterpart $\so_q(4)$. To this end, we replace the classical generators of the $\su(2)$ gauge algebra with the corresponding quantum operators, using the Loop Quantum Gravity (LQG) techniques \cite{lqg}. This construction relies on earlier results \cite{DP2}, \cite{DP3}. Having defined the quantum constraints we turn to calculating their commutators. It turns out that in general the algebra of commutators of constraints is anomalous, but the anomaly can be removed when a natural condition is imposed (\ref{q-dim}). It is remarkable that the primary reason for the emergence of the deformed Hopf structure in the theory is anomaly cancellation.

The analysis of Sect.\ 4 reveals not only the fact that the algebra of quantum constraints is deformed (which by itself is not very informative, because we can always rescale the quantum constraints, so as to make the algebra undeformed), but the form of the $R$-matrix, which carries  rescaling-independent information about the Hopf algebra structure. We discuss this $R$-matrix in detail in Sect.\ 5.

In general knowing just the algebra and the $R$-matrix  is not sufficient to reconstruct the whole Hopf algebra structure. In the case at hand this can be done, fortunately. The reason is that in the case of $\mathfrak{sl}(2,\mathbb{C})$ (and of its real forms such as $\su(2)$) the complete classification of possible deformations is known, and knowing the $R$-matrix one can read-off the form of the coproduct and the antipode.

In Sect.\ 2 we are starting with the  classical symmetry algebra $\so(4)$ which we decompose into two copies of $\su(2)$. After quantization each copy becomes a deformed Hopf algebra $\su_q(2)$. As we show in Sect.\ 6 these two deformed groups can be combined into a Hopf algebra of deformed $\so(4)$.
Having obtained the deformed symmetry algebra for the case of the Euclidean de Sitter quantum gravity in 3D, we now want to make the contraction $\Lambda \rightarrow 0$, so as to obtain a symmetry that replaces, according to what was said above, the standard Poincar\'e symmetry in the case of quantum space-time. It turns out that taking the contraction limit is highly nontrivial and can be performed if the deformation parameters $q_1,q_2$ of the two deformed $\su_q(2)$ algebras satisfy $q_1 = q_2^{-1}$ \cite{Celeghini:1990xx} (the divergencies are not visible on the level of the algebra, but can be seen when one wants to find the expression of coproducts in the contraction limit; for this reason they were missed  in\cite{Amelino}.) The so constructed symmetry algebra of flat quantum space-time in 3D turns out to be the $\kappa$-Poincar\'e algebra. This is the main and final result of our paper, and it is presented in Sect.~7.

We comment on the possible physical implications of our analysis in the presence of coupled point particles in Sect.~8.
Conclusions are presented in Sect.~9.

\section{Classical phase space and  constraint algebra}\la{Algebra-c}

We want to study the algebra of constraints of Euclidean gravity on a 3-dimensional manifold $\sM$ with positive cosmological constant $\Lambda$. The phase space of the theory is parametrized by 3-dimensional connection 1-form $\omega$ and the $\su(2)$-valued triad 1-form $e$.
The action of the theory is given by
\be\la{action}
S[e,\omega]=\kappa\int_{\sM}\tr[e\wedge
F(\omega)+\frac{\Lambda}{3} e \wedge e\wedge e]\,,
\ee
where $F(\omega)$ is the connection curvature, the trace is defined through an $\su(2)$ Killing form and $\kappa=(4\pi G)^{-1}$  is the Planck mass. Upon a canonical decomposition of $\sM$ into a time direction and a Riemann surface $\Sigma$, namely $\sM= \Sigma \times \R$, the canonical phase space is parametrized by the pull back of $\omega$ on $\Sigma$, which we denote $A^i_a = 1/2 \epsilon^i_{\, jk}\omega^{jk}_a$,  and its conjugate momentum $E^b_j = \kappa\epsilon^{bc} e^k_c
\eta_{jk}$. In our notation, $a = 1, 2$ are space coordinate indices on $\Sigma$, $i, j
= 1, 2, 3$ label internal $\su(2)$ indices, which we raise and lower with the Killing metric $\delta_{ij}$, and $\epsilon^{ab}=-\epsilon^{ba}$ with $\epsilon^{12}=1$.
The canonical phase space variables satisfy the Poisson bracket
\be\la{symplectic}
\{A^i_a(x), E^b_j(y)\}=\delta^b_a \delta^i_j \delta^{(2)}(x,y)~.
\ee
The variation of the action with respect to the variables $e_t^i$ and $\omega_t^{ij}$ leads to
two sets of smeared constraints
\ba
&&G[\alpha]=\int_\Sigma \alpha^i G_i=\int_\Sigma \alpha^i D_A E_i=0\,,\label{Gauss}\\
&&C_{\va \Lambda}[N]=\int_\Sigma N_i C^i_{\va \Lambda}=\int_\Sigma N_i( \kappa F^i(A)+\frac{\Lambda}{2\kappa} \epsilon^{ijk}E_j \wedge E_k)=0\,,\la{Curvature}
\ea
where $\alpha, N$ are arbitrary $\su(2)$-valued test functions, independent of the connection and momentum variables.
The constraint \eqref{Gauss} is called the Gauss constraint and it implements the local $SU(2)$ gauge invariance of the theory; the second constraint \eqref{Curvature} is called the curvature constraint and it encodes the information that the connection is no longer flat (as it was in the $\Lambda=0$ case) and it also generates gauge symmetries.

The classical constraint algebra of the theory reads
\ba\label{cl-algebra}
\{C_{\va \Lambda}[N],C_{\va \Lambda}[M]\}&=& \Lambda \ G[[N,M]]\,,\n\\
\{C_{\va \Lambda}[N],G[\alpha]\}&=&C_{\va \Lambda}[[N,\alpha]]\,,\n\\
\{G[\alpha],G[\beta]\}&=&G[[\alpha,\beta]]\,,
\ea
where $[a,b]^i=\epsilon^{i}_{\ jk} a^jb^k$ is the commutator of $\su(2)$.

In order to use the techniques and results of \cite{DP2, DP3}, we define a new non-commutative connection
\be\la{conn}
A^{\pm i}_a=A^i_a\pm\sqrt{\Lambda} e^i_a=A^i_a\pm\frac{\sqrt{\Lambda}}{\kappa}\epsilon_{ba}E^b_i\,,
\ee
such that the Gauss and curvature constraints can be expressed as
\ba\la{curvature2}
&&C_{\va \Lambda}[N]=\frac{1}{2}\left(H^+[N]+H^-[N]\right)\,, \\
&&G[N]=\frac{1}{2\sqrt{\Lambda}}\left(H^+[N]-H^-[N]\right)\,,
\ea
where
\be\la{curvaturepm}
H^\pm[N]\equiv \kappa \int_\Sigma N_i F^i(A^\pm)
\ee
is the curvature constraint for the non-commutative connection $A_{\va \Lambda}$.

Therefore, the set of constraints \eqref{curvaturepm} is equivalent to the constraints \eqref{Gauss}, \eqref{Curvature}, and their algebra is
\begin{equation}\label{HH}
\begin{split}
\{H^\pm[N],H^\pm[M]\} & =  \pm2\sqrt{\Lambda} \ H^\pm[[N,M]]\\
\{H^+[N],H^-[M]\} & = 0\,,
\end{split}
\end{equation} 
corresponding to two copies of $\su(2)$, i.e. the constraints of the theory generate a local $\su(2)\oplus\su(2)$ symmetry.

\section{Constraints and symmetries of maximally symmetric space-times}\la{symmetries}

We want to show now the relation between the structures introduced in the previous section and the diffeomorphism constraints, and how, for a maximally symmetric space-time, these reproduce the algebra of relativistic symmetries.

The diffeomorphism constraints, generating the transformation $x^\mu\rightarrow x^\mu+\xi^\mu$, can be expressed as a linear combination of the Gauss and curvature constraints by means of smearing functions depending on phase space variables:
\begin{equation}
D[\xi]=C_\Lambda[\xi^\mu\,e^i_\mu]+G[\xi^\mu\,A_\mu^i]\,.\label{diff}
\end{equation}
This is proved in Appendix~\ref{appendix}, where we derive  the  Poisson brackets action of the diffeomorphism constraints on the phase space variables and show that it is the Lie derivative along the vector field $\xi^\mu$.
Substituting the relations \eqref{conn} into \eqref{diff}, the diffeomorphism constraints can be expressed in terms of the $H^\pm$ constraints \eqref{curvaturepm} as
\be
D[\xi]=\frac{1}{2\sqrt{\Lambda}}\,\left(H^+[\xi^\mu\,A_\mu^+]-H^-[\xi^\mu\,A_\mu^-]\right)\,,
\label{diff2}
\ee
where it is important to notice that the smearing functions depend in general on the phase space variables.

We restrict now to the case of a maximally symmetric space-time, and in particular we consider the case of an empty homogeneous and isotropic universe with positive cosmological constant $\Lambda$ in 2+1 dimensions and Euclidean signature, which we call `Euclidean de Sitter' and whose manifold can be described by the 3-sphere $S^3$ with (constant) radius $1/\sqrt\Lambda$, invariant under SO(4) symmetries.
Being a maximally symmetric space-time, it admits 6 Killing vectors, forming the $\mathfrak{so}(4)$ algebra.
In order to make contact with relativistic symmetries, we express the $\mathfrak{so}(4)$ Killing vectors as generators of time translations, space translations, boosts and rotations for the Euclidean (2+1)D case, which we call respectively $\xi_{(E)}$, $\xi_{(P_a)}$, $\xi_{(N_a)}$ and $\xi_{(M)}$.  Here we denote  $\xi_{(X)}$  the Killing vector labelled by $X$. These vectors satisfy the commutation relations
\begin{equation}\label{SO4killing}
\begin{gathered}\left[\xi_{(E)},\xi_{(P_{a})}\right]=\Lambda\xi_{(N_{a})},\qquad\left[\xi_{(P_{1})},\xi_{(P_{2})}\right]=\Lambda\xi_{(M)}\\
\left[\xi_{(N_{a})},\xi_{(E)}\right]=\xi_{(P_{a})},\qquad\left[\xi_{(N_{a})},\xi_{(P_{b})}\right]=-\delta_{ab}\xi_{(E)},\qquad\left[\xi_{(N_{1})},\xi_{(N_{2})}\right]=\xi_{(M)},\\
\left[\xi_{(M)},\xi_{(N_{a})}\right]=\epsilon_{ab}\,\xi_{(N_{b})},\qquad\left[\xi_{(M)},\xi_{(P_{a})}\right]=\epsilon_{ab}\,\xi_{(P_{b})},\qquad\left[\xi_{(M)},\xi_{(E)}\right]=0,
\end{gathered}
\end{equation}
where for $\xi_{(I)} = \xi_{(I)}^\mu\partial_\mu$, and $[\xi_{(I)},\xi_{(J)}]=\xi_{(I)}^\mu\partial_\mu \xi_{(J)} - \xi_{(J)}^\mu\partial_\mu \xi_{(I)} $.

We now notice that the symmetries of the space-time manifold can be conveniently described in terms of left and right invariant vector fields of the $SU(2)$ group \cite{SU(2)Avramidi}, $L_i$ and $R_i$, respectively. In fact, the Killing vectors~\eqref{SO4killing}, which are elements of the $\so(4)$ algebra, can be arranged as two commuting sets of $\su(2)$ Killing vectors $\xi^+_{(i)}$ and $\xi^-_{(i)}$ through the relations
\begin{equation}\label{KillingSO4SU2}
\begin{gathered}\xi_{(E)}=-\frac{\sqrt{\Lambda}}{2}\left(\xi_{(3)}^{+}+\xi_{(3)}^{-}\right),\qquad\xi_{(P_{a})}=-\frac{\sqrt{\Lambda}}{2}\epsilon_{ab}\left(\xi_{(b)}^{+}-\xi_{(b)}^{-}\right),\\
\xi_{(M)}=\frac{1}{2}\left(\xi_{(3)}^{+}-\xi_{(3)}^{-}\right),\qquad\xi_{(N_{a})}=-\frac{1}{2}\left(\xi_{(a)}^{+}+\xi_{(a)}^{-}\right),
\end{gathered}
\end{equation}
from which it follows that $\xi^\pm_{(i)}$ satisfy 2 copies of the $\mathfrak{su}(2)$ algebra (in the ``chiral'' basis)
\begin{equation}
[\xi_{(i)}^\pm,\xi_{(j)}^\pm]=\pm2\epsilon_{ij}^{\ \ k}\xi^\pm_{(k)},\qquad  [\xi_{(i)}^+,\xi_{(j)}^-]=0
\end{equation}
so that  $\xi^+_{(i)}\equiv L_{(i)}$ and $\xi^-_{(i)}\equiv  R_{(i)}$. This enables one to choose as triads (modulo a factor $1/\sqrt{\Lambda}$) the inverse co-triads $X^i$, $Y^i$ corresponding to $L_{(i)}$ and $R_{(i)}$, respectively. For instance, in the former case, one has
\be
e^i=\frac{1}{\sqrt{\Lambda}}\, X^i\qquad X^i_{\mu} \,L^\mu_{(j)}=\delta^i_j\,,
\ee
such that spin connections are given by
\be
\omega^{ij}=\epsilon^{ij}_{\phantom{12}k} \,X^k\,,
\ee
and $A^{- i}=0$, $A^{+i}=2\, X^i$. Hence, one finds
\be\label{DH+}
D[\xi^+_i]=D[L_{(i)}]=\frac{1}{\sqrt{\Lambda}}\,H^+[L^\mu_{(i)}\,X_\mu]=\frac{1}{\sqrt{\Lambda}}\,H^+[\delta_{(i)}]
\ee
with $\delta_{(i)}^j=\delta^j_i$. Similarly, by choosing $e^i=Y^i/\sqrt{\Lambda}$, with $Y^i_{\mu} \,R^\mu_{(i)}=\delta^i_j$, one gets $\omega^{ij}=-\frac{1}{2}\,\epsilon^{ij}_{\phantom{12}k} \,Y^i$, from which $A^{+ i}=0$, $A^{-i}=-2\, Y^i$ and
\be\label{DH-}
D[\xi^-_i]=D[R_{(i)}]=\frac{1}{\sqrt{\Lambda}}\,H^-[\delta_{(i)}]\,.
\ee

This shows that when restricting to $S^3$, the diffeomorphism constraints can be expressed in terms of smearing functions that do not depend on the phase variables, and in particular they split into two sets of constraints corresponding to the two copies of constraints (\ref{curvaturepm}).
Therefore, in order to study symmetries of (Euclidean) deSitter space-time it is enough to consider the algebra \eqref{HH}, in which the smearing functions are proportional to $\delta$'s.
Let us see this in more detail.

Let 
\begin{equation}\label{LR}
{\cal L}_i = D[\xi^+] = H^+[\delta_{(i)}]/\sqrt{\Lambda}, \qquad {\cal R}_i = D[\xi^-]=H^-[\delta_{(i)}]/\sqrt{\Lambda}
\end{equation}
be the generators associated to the Killing vectors $\xi^\pm_i$.
Using Eq.~\eqref{HH} we find
\begin{equation}
\begin{split}\left[{\cal L}_{i},{\cal L}_{j}\right]= & \left[D\left[\xi_{i}^{+}\right],D\left[\xi_{j}^{+}\right]\right]=\frac{1}{\Lambda}\left[H^{+}\left[\delta_{(i)}\right],H^{+}\left[\delta_{(j)}\right]\right]\\
= & \frac{2}{\sqrt{\Lambda}}H^{+}\left[\left[\delta_{(i)},\delta_{(i)}\right]\right]
=  \epsilon_{ij}{}^{ k}\frac{2}{\sqrt{\Lambda}}H^{+}\left[\delta_{(k)}\right]=2\epsilon_{ij}^{\ \ k}{\cal L}_{k}.
\end{split}
\end{equation}
Similarly, using again \eqref{HH}, we find that ${\cal L}_i$ and ${\cal R}_i$ form two copies of $\mathfrak{su}(2)$
\begin{equation}\label{SU2SU2}
[{\cal L}_i,{\cal L}_j] = 2\epsilon_{ij}^{\ \ k}{\cal L}_k \qquad
[{\cal R}_i,{\cal R}_j] = -2\epsilon_{ij}^{\ \ k}{\cal R}_k, \qquad [{\cal L}_i,{\cal R}_j] = 0,
\end{equation}
i.e. ${\cal L}_i$ and ${\cal R}_i$ are left and right generators of $\su(2)$ in the chiral bais.

We define now the generators associated to the Killing vectors~(\ref{SO4killing}) as
\begin{equation}
E=D[\xi_{(E)}],\qquad P_a=D[\xi_{(P_a)}],\qquad N_a=D[\xi_{(N_a)}],\qquad M=D[\xi_{(M)}].
\end{equation}
From the definitions~\eqref{LR} and the maps~\eqref{KillingSO4SU2} it follows for instance that
\begin{equation}
E = \frac{\sqrt{\Lambda}}{2}D\left[\xi_{(3)}^{+}+\xi_{(3)}^{-}\right] =
\frac{\sqrt{\Lambda}}{2}\left(D\left[\xi_{(3)}^{+}\right] + D\left[ \xi_{(3)}^{-}\right] \right) = \frac{\sqrt{\Lambda}}{2}\left({\cal L}_3 + {\cal R}_3\right).
\end{equation}
Then, using~\eqref{LR},~\eqref{KillingSO4SU2} and~\eqref{SU2SU2}, we obtain the algebra
\begin{equation}\label{SO4}
\begin{gathered}\left[E,P_{a}\right]=\Lambda N_{a},\qquad\left[P_{1},P_{2}\right]=\Lambda M\\
\left[N_{a},E\right]=P_{a},\qquad\left[N_{a},P_{b}\right]=-\delta_{ab}E,\qquad\left[N_{1},N_{2}\right]=M,\\
\left[M,N_{a}\right]=\epsilon_{ab}\,N_{b},\qquad\left[M,P_{a}\right]=\epsilon_{ab}\,P_{b},\qquad\left[M,E\right]=0.
\end{gathered}
\end{equation}
This algebra is isomorphic to the algebra of the associated Killing vectors~(\ref{SO4killing}), showing the consistency of our approach, and moreover one notices that this is nothing but the $\mathfrak{so}(4)$ algebra expressed in terms of translation, boost and rotation generators for the `Euclidean' de Sitter space-time (see for instance~\cite{caccia}), as can be easily shown by identifying $E=\sqrt{\Lambda}M_{03}$, $P_a=\sqrt{\Lambda}M_{a3}$, $N_a=M_{0a}$, $M=M_{12}$, where
\begin{equation}
\left[M_{AB},M_{CD}\right]=\delta_{[A}^{e}\delta_{[B}^{f}\eta_{C]D]}M_{ef}=\eta_{AC}M_{BD}+\eta_{BD}M_{AC}-\eta_{AD}M_{BC}-\eta_{BC}M_{AD}\,,
\end{equation}
and $\eta=\text{diag}\left(++++\right)$, with $A=0,1,2,3$.

It is worth noting that the (2+1)D de Sitter algebra (with Lorentzian signature) can be obtained by a Wick rotation for boosts and time translation: $E\rightarrow -iE$, $N_a\rightarrow -iN_a$.

\section{Quantum phase space and  constraint algebra}\la{Algebra-q}

In order to quantize the theory presented in the previous section, we are going to use the LQG formalism.
In three dimensions, the Riemannian theory with vanishing cosmological constant can be quantized using LQG techniques both in the covariant and canonical formalisms and the two quantizations  have been shown to be equivalent \cite{NP} (see also \cite{AGN} for a review of this topic). In the case of $\Lambda\neq0$ the canonical quantization has been implemented in \cite{DP2, DP3}, and it was shown to reproduce the physical transition amplitudes of the Turaev-Viro state sum \cite{TV}, which provides a covariant quantization of the theory (see also \cite{Girelli, Geiller} for alternative approaches to the LQG quantization). Here we want to study the off-shell algebra of the constraints, which is the new result of this section. Let us start by briefly reviewing the main ingredients of the LQG quantization (see \cite{lqg} for more details).

The auxiliary kinemtical Hilbert
space $\Hk$ of the theory is constructed by replacing functionals of the connection variable with cylindrical functionals of holonomies along paths $\gamma \subset
\Sigma$, which are called generalized connections. Holonomies are given by a path ordered exponential of $A$, namely
\be\label{hol}
h_{\gamma}[A]=P \exp\int_{\gamma} A \,,
\ee
and they represent quantum, polymer-like excitation of the gravitational field.
A particular example of gauge-invariant cylindrical functionals of holonomies are represented by spin networks. Given a finite graph $\Gamma \subset \Sigma$, whose links and nodes we indicate respectively $\ell, n$, we assign to each link with a spin $j_\ell$ labelling an $SU(2)$ irreducible representation and to each node an invariant tensor $\iota_n$, called an intertwiner, in the tensor product of $SU(2)$ irreducible representations labelling the edges attached to the node. The corresponding spin network functional is defined
\be
\Psi_{\gamma, \{\iota_n\} \{j_\ell\}}[A]=\bigotimes_{n\subset\Gamma} \iota_n\bigotimes_{\ell\subset n}\stackrel{j_\ell}{\Pi}(h_{\ell}[A])\,,
\ee
where we have omitted the indices of the representation matrices, $\Pi$'s, and of the intertwiners to lighten the notation. Notice that, the assignment of an $SU(2)$ irreducible representation to each link implies that there is a certain vector space associated to each link, namely the representation vector space; for a given spin-$j$ representation, the associated vector space is the tensor product of $2j$ copies of $\C^2$, where $\C^2$ is the representation vector space of the fundamental representation spin-$1/2$.
By introducing the Ashtekar-Lewandowski measure \cite{AL} (constructed in terms of the $SU(2)$ Haar measure) in the space of generalized connections in order to define a notion of kinematical inner product, it can be shown that spin network functions define a complete orthogonal basis of $\Hk$.

In order to define a representation of the action of the Gauss and curvature constraints on $\Hk$, we start with the action of quantum holonomies and fluxes.

The generalized connection is represented as a self-adjoint quantum holonomy operator in the kinematical Hilbert space that acts  by multiplication
\be
{\hat h_\gamma[A]} \Psi[A] \; = \; h_\gamma[A] \Psi[A]\,.
\label{ggcc}
\ee

The operator associated to the momentum $E$ in $\Hk$ is defined by smearing the 1-form $E^a_i\epsilon_{ab}$ along the one-dimensional path $\eta^a(t)\subset \Sigma$, namely
\be
E(\eta)\equiv \int E^{a}_i\tau^i\epsilon_{ab}\frac{d\eta^a}{dt}= \int E^{a}_i\tau^i n_a dt\,,
\ee
where we have defined $n_a\equiv
\epsilon_{ab}\frac{d\eta^a}{dt}$ the normal to the path. Therefore, $E(\eta)$ represents 
 the flux of the $E$ field across one-dimensional line $\eta$. By replacing $E^{a}_i$ with the functional derivative with respect to $A^i_a$,
the action of the associated quantum operator  on a spin network link $\gamma$ can be defined as
\ba \label{flux}\hat E(\eta)\Psi_{\gamma}[A]=\frac{1}{2}\hbar
\left\{\begin{array}{ccc} \!\! o(p) \tau_i \Psi_{\gamma}[A]\ \ \mbox{if $\gamma$
ends at $\eta$}\\ o(p) \Psi_{\gamma}[A]\tau_i\ \ \mbox{if $\gamma$ starts at
$\eta$}\end{array}\right.,
\ea
 where $o(p)=\pm1$ is the orientation of
the intersection $p\in \Sigma$ between the two oriented curves $(\eta,\gamma)$.

Due to the smeared nature of the quantum phase space variables, in order to write the quantum version of the constraints  \eqref{Gauss}, \eqref{curvature2}, we need first to introduce a regulator consisting of  an arbitrary finite cellular decomposition $\Delta_{\Sigma}$
of $\Sigma$---with plaquettes $p\in\Delta_{\Sigma}$ of coordinate area smaller or equal to $\epsilon^{2}$---. The constraints can then be written as
\ba
&&H^\pm\left[N\right]=\lim_{\epsilon\rightarrow0}\sum_{p\in\Delta_{\Sigma}}\tr\left[N_{p}\,
W_{p}\left(A^\pm\right)\right]=0\,,\la{p-Curvature}
\ea
where $W_{p}(A^\pm)$ is the holonomy of the connection $A^\pm$ around the given plaquette $p$.

The quantization of the holonomy of the non-commutative connection as an operator acting on the kinematical Hilbert space of LQG has been performed in \cite{DP2}. In that analysis, one considers two holonomies, $h_{\eta}, h_{\gamma}$, of the non-commutative connection on intersecting paths $\eta, \gamma$ acting on the vacuum. The action of the first holonomy is the same as for the case of the commutative holonomy \eqref{ggcc} and it simply creates a spin network state associated to the path defining the holonomy. The second non-commutative holonomy then acts non-trivially on this state. One can expand it in powers of $\sqrt{\Lambda}$ and look at this action order by order. In this expansion, the presence of powers of flux operators all acting at the same point introduces ordering ambiguities. Relying on the use of the Duflo map to solve these ambiguities, it was shown that the series expansion converges and the action can be expressed in terms of
 the Kauffman bracket \cite{KL} crossing identities, namely
 \ba
&&\hat h_{\eta}\left(A^+\right)\triangleright
\hat h_{\gamma}\left(A^+\right)\,\left|0\right\rangle
=
\begin{array}{c}\psfrag{a}{$\eta$}
\psfrag{b}{$\gamma$}
\includegraphics[width=1.cm]{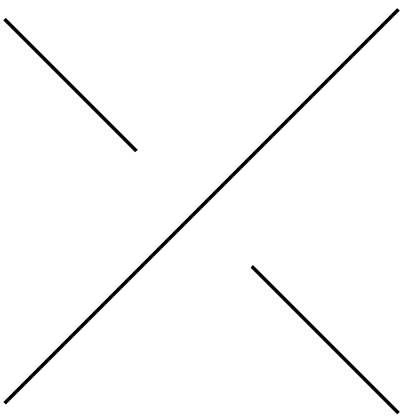}\end{array}=A\begin{array}{c}
\includegraphics[width=1.cm]{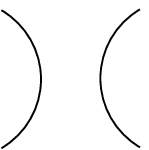}\end{array}+A^{-1}  \begin{array}{c}
\includegraphics[width=1.cm]{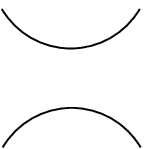}\end{array}\la{cross2a}\,,\\
&&\hat h_{\eta}\left(A^-\right)\triangleright
\hat h_{\gamma}\left(A^-\right)\,\left|0\right\rangle
=
\begin{array}{c}
\includegraphics[width=1.cm]{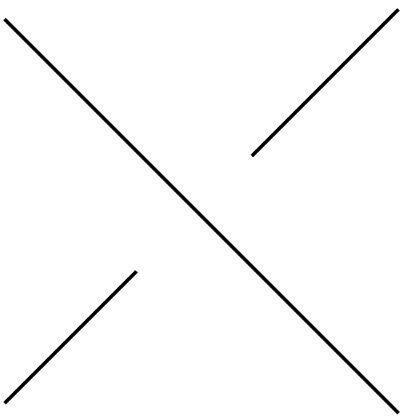}\end{array}=A^{-1}\begin{array}{c}
\includegraphics[width=1.cm]{qL1.eps}\end{array}
+A  \begin{array}{c}
\includegraphics[width=1.cm]{qL2.eps}\end{array}\la{cross2b}\,.
\ea
In this case, the over-crossing and under-crossing notation is used to distinguish between the two connections $A^+$ and $A^-$. The skein relations associated to the two connections are simply related by the switch of deformation parameters $A\rightarrow A^{-1}$ (this relation will play a crucial role when performing the contraction of the co-prodruct sector).

Using this result, the algebra of the quantum constraint $\hat H^+\left[N\right]$ with itself (it is immediate to see that the same result applies also to $\hat H^-\left[N\right]$)  on a gauge invariant state has been computed in \cite{DP3}, by means of techniques developed in \cite{DP1, DP4}.It was found that the algebra  is anomaly-free if and only if the following condition holds
\be\la{q-dim}
\tr[W_{p}]=\begin{array}{c}\includegraphics[width=1.cm]{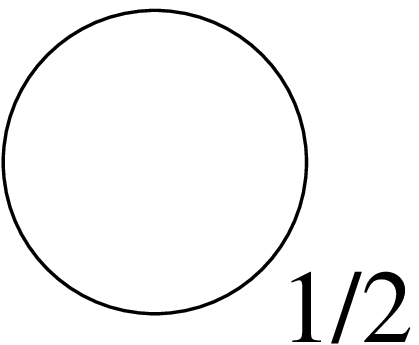}\end{array}=-(A^2+A^{-2})\,,
\ee
where $A=e^{{i \hbar \sqrt{\Lambda}}/{4\kappa} }$ and $W_{p}$ is the holonomy of the commutative connection $A^i$ around $p$. With this condition, corresponding to the second ({\it quantum dimension}) Kauffman bracket, the off-shell algebra of the generators $\hat H^\pm\left[N\right]$ can be computed by acting on a simple not gauge-invariant state living on the dual cellular decomposition  $\Delta_{\Sigma^*}$ formed by plaquettes $p^*$s dual to the $p$s. More precisely, given a plaquette $p$ on which the constraints are defined, we consider the state schematically depicted
\be
|\Psi\rangle=
\begin{array}{c}
\includegraphics[width=3.cm]{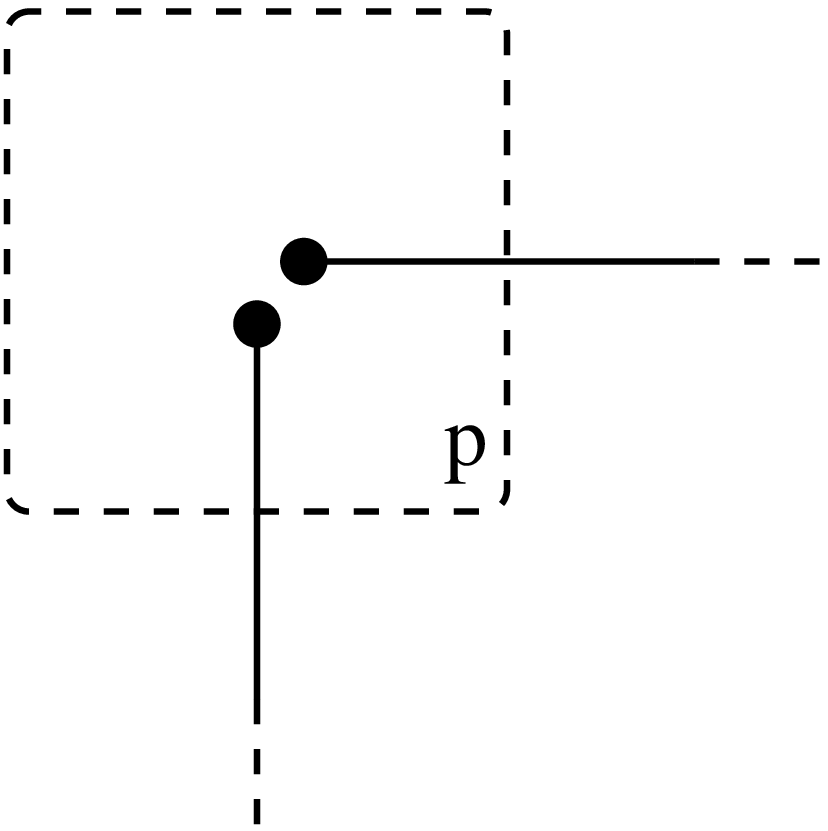}\,,
\end{array}
\ee
and formed by two open links of spin-1/2 obtained by removing a bivalent intertwiner from a single link living in  $\Delta_{\Sigma^*}$. Generalization to more open links and/or higher spin is a lengthy but straightforward calculation that does not affect the form of the algebra.

It is immediate to see that commutators between generators defined on different plaquettes commute, hence we can restrict to the case of $\hat H^\pm[N_p]$ at the same plaquette, when studying the algebra.
By repeated applications of the skein relations \eqref{cross2a}, \eqref{cross2b}, \eqref{q-dim}, a lengthy but straightforward calculation yields (upon removal of the anomalous terms by means of \eqref{q-dim}) the off-shell algebra
\ba
\left[\hat H^\pm[N_p],\hat H^\pm[M_p]\right]|\Psi\rangle&=&\pm\sqrt{\Lambda} \left(A^2+A^{-2}\right)\ \hat H^\pm[[N_p,M_p]]|\Psi\rangle\,,\la{HH-q}\\
\left[\hat H^\pm[N_p],\hat H^\mp[M_p]\right]|\Psi\rangle&=&0\,.\la{HH-q2}
\ea
In the  (naive) classical limit $\hbar\rightarrow0$ with $\Lambda$ and $\kappa$ kept finite, $A^2+A^{-2}\rightarrow 2$, the algebra gives back the first Poisson brackets  in \eqref{HH}\,.

Therefore, we see from \eqref{HH-q} that at the quantum level the algebra of constraints is deformed. We know already from the $q$-deformed skein relations  \eqref{cross2a}, \eqref{cross2b}, \eqref{q-dim} that the new symmetry replacing the classical $SU(2)$ one, is described by the quantum group (deformed Hopf algebra) which we denote by\footnote{It is common use in the literature to describe the resulting quantum group in terms of a generic real form $U_q(\mathfrak{sl}(2,\mathbb{C})_\mathbb{R})$. The choice of real form is determined by the reality conditions on the algebra (i.e. the choice of the $*$-structure), giving rise to three different possibilities: $U_q(\mathfrak{su}(2))$, $U_q(\mathfrak{su}(1,1))$ and $U_q(\mathfrak{sl}(2,\mathbb{R}))$ (see for instance~\cite{MajidFound}). In our case the Hermiticity of the symmetry generators is defined, corresponding to the one of $\su(2)$ (as  may be obvious since we start our analysis from $SU(2)$ symmetries), so that we can specify from the beginning which real form characterizes our deformation. We will come back to this point at the end of Sec.~\ref{fromSO2toSO4}, where we will discuss the reality conditions of our deformed symmetry algebra.} $\su_q(2) = U_q (\mathfrak{su}(2))$.
This is a well known fact, but to clarify the investigation of deformed space-time symmetry arising from the discrete structures underlying the loop quantization, it is important to make this point clear. In order to do that, the algebra itself is not very relevant, since one can always rescale the constraint operators so as to formally remove the deformation. The relevant structure to look at is the $R$-matrix structure behind the crossing  properties of two non-commutative holonomies defining the constraints. By explicitly showing that the $q$-deformed crossing identity \eqref{cross2a}, \eqref{cross2b} can be represented in terms of the $\su_q(2)$ $R$-matrix, we can unravel the quantum group symmetry encoded in the constraint algebra of 2+1 LQG with positive cosmological constant. We do this in the next section.

\section{The $R$-matrix}
\label{sec:$R$-matrix}

As the constraint algebra encodes the isometries of space-time and the constraints are expressed in terms of non-commutative holonomies, the crossing properties of such non-commutative holonomies are expected to encode the information about the $R$-matrix of the quasi-triangular bialgebra governing the symmetries of quantum space-time, and eventually the braiding properties of point particles coupled to this quantum background geometry (see Section \ref{Particles} below for a discussion on this). In this section we are going to unravel such connection.
 Namely, we want to study the $R$-matrix associated to the crossing of two non-commutative holonomies which are part of the  loops operators corresponding to the generators $\hat H^+[N_p]$.

Let us concentrate on the constraint $\hat H^{+}$, a completely similar derivation follows for $H^-$.
We have seen that non-commutative holonomy operators satisfy \eqref{cross2a},
which also implies
\be\la{q-dim2}
\begin{array}{c}\includegraphics[width=1.cm]{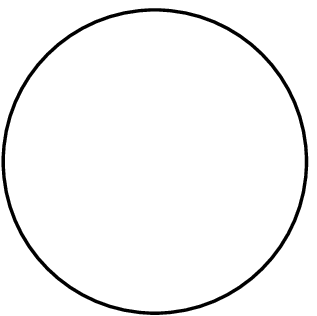}\end{array}=-(A^2+A^{-2})\,,
\ee
in order for the algebra of the constraints to close.

Let us consider two holonomies of the non-commutative connection \eqref{conn} on crossing loops like in the figure below (corresponding to the Hopf link)
\bee\la{Rmatrix}
\begin{array}{c}\includegraphics[width=5.cm]{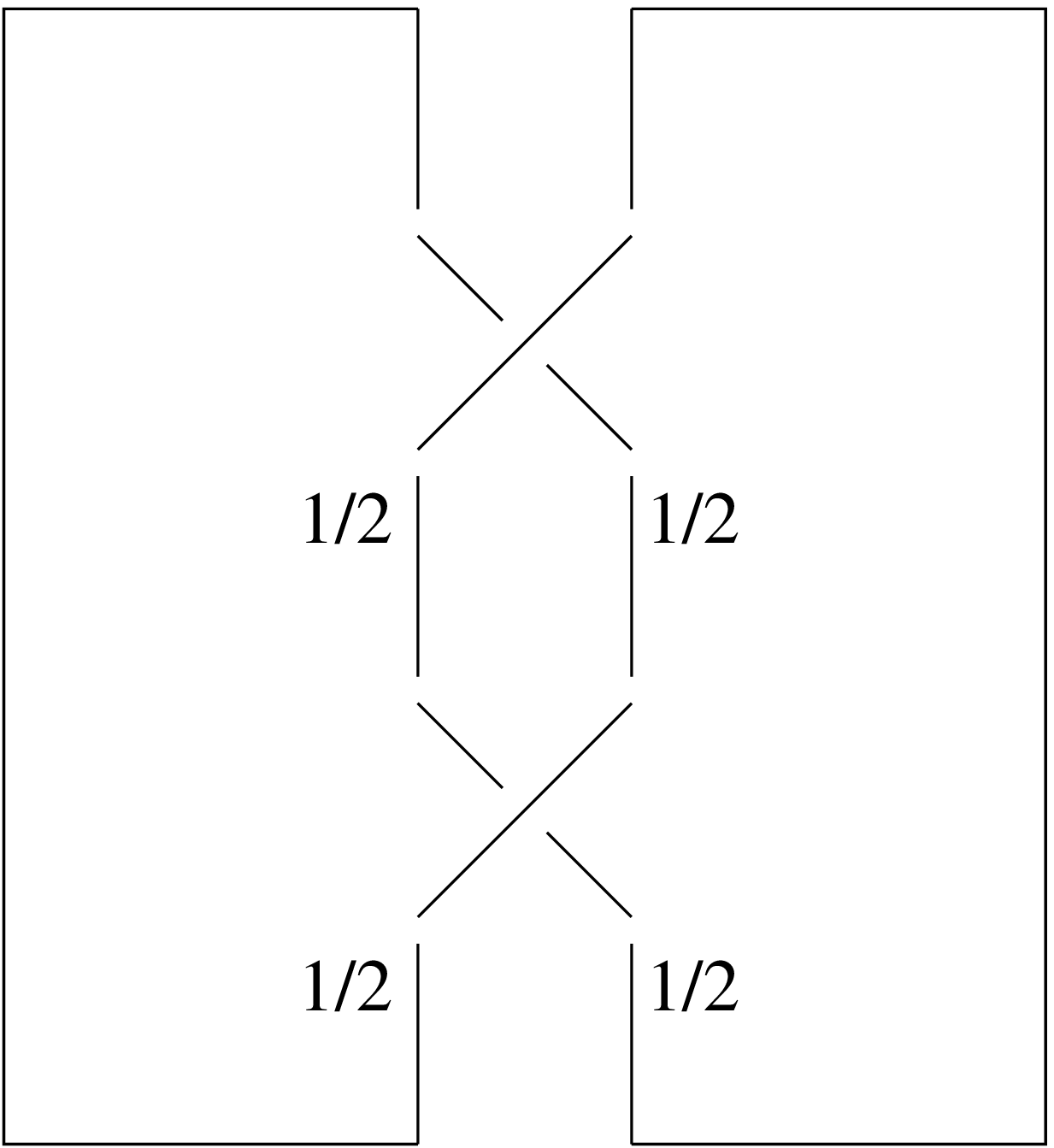}\end{array}\,.
\eee
We see that each crossing is followed by two non-overlapping parts. When acting on the vacuum, these straight (in the figure) parts of the holonomies act like standard commutative holonomy operators, thus creating two single link states in the fundamental representation $1/2$. We can then interpret each crossing of the two holonomies as operators acting on these two states. Since each link-1/2 carries a representation vector space $V=\C^2$, we can derive the form of the $R$-matrix associated to the generators $\hat H^+[N_p]$ by studying the action of the crossing operators on the tensor product vector space $V\otimes V$.
We can then  interpret the diagrammatic relations \eqref{cross2a}, \eqref{q-dim2} above as relations between operators:
\ba
&&\begin{array}{c}\includegraphics[width=1.cm]{x-quantum.eps}\end{array}: V\otimes V~\rightarrow ~V\otimes V\n\\
&&\begin{array}{c}\includegraphics[width=1.cm]{x-quantum-2.eps}\end{array}: V\otimes V~\rightarrow~ V\otimes V\n\\
&&\begin{array}{c}\includegraphics[width=1.cm]{qL1.eps}\end{array}: V\otimes V~\rightarrow~ V\otimes V\n\\
&&\begin{array}{c}\includegraphics[width=1.cm]{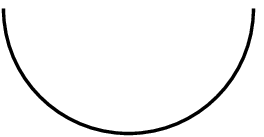}\end{array}: V\otimes V~\rightarrow ~\C\n\\
&&\begin{array}{c}\includegraphics[width=1.cm]{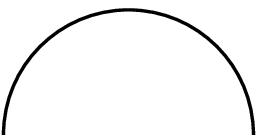}\end{array}: \C~\rightarrow~  V\otimes V\,.
\ea
Given an orthonormal basis of $V=\C^2$ formed by the vectors $v_1, v_2$, we want to define the action of the cup and cap operators, respectively $\begin{array}{c}\includegraphics[width=.7cm]{cup.eps}~ {\rm and}~ \includegraphics[width=.7cm]{cap.eps}\end{array}$, on such basis which is compatible with the relations \eqref{cross2a} and such that the bracket \eqref{q-dim2} as well as the identity
\be\la{Id}
\begin{array}{c}\includegraphics[width=5.cm]{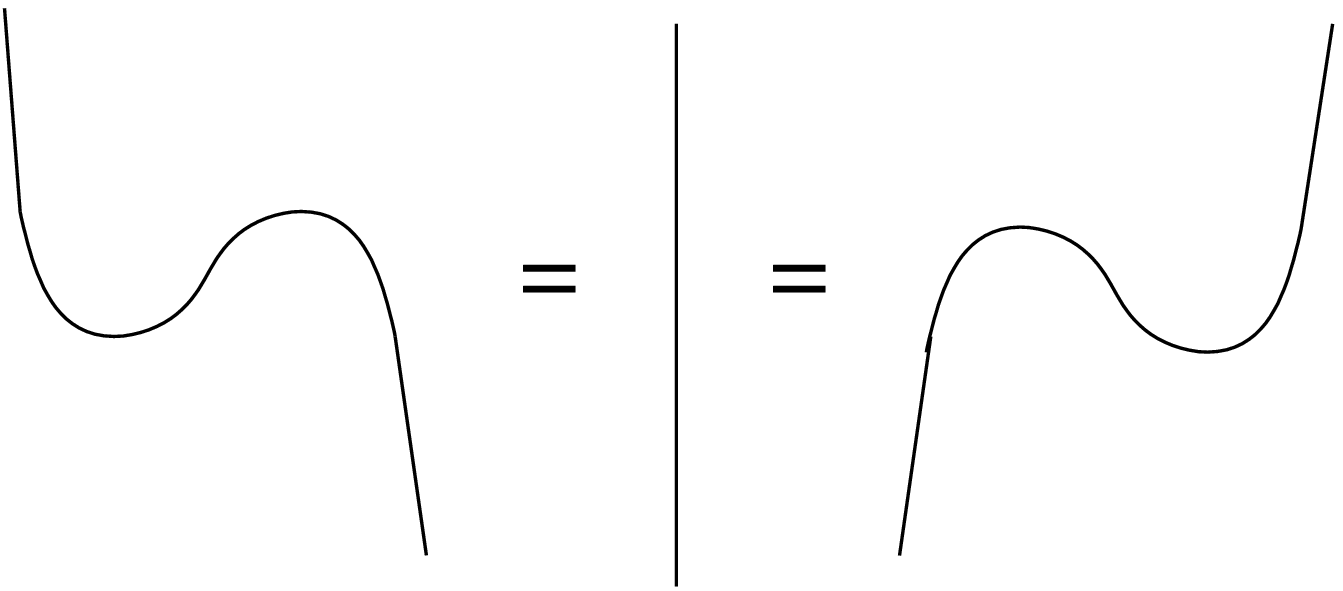}\end{array}
\ee
are satisfied. This happens for the following actions of the cup and cap operators:
\ba
\begin{array}{c}\includegraphics[width=1.cm]{cup.eps}\end{array}&:& V\otimes V~\rightarrow ~\C\n\\
&&v_1\otimes v_1~\rightarrow ~0\n\\
&&v_1\otimes v_2~\rightarrow ~A\n\\
&&v_2\otimes v_1~\rightarrow ~-A^{-1}\n\\
&&v_2\otimes v_2~\rightarrow ~0\la{cup}\\
\n\\
\begin{array}{c}\includegraphics[width=1.cm]{cap.eps}\end{array}&:& \C~\rightarrow~  V\otimes V\n\\
&&1~\rightarrow ~-A v_1\otimes v_2+A^{-1}v_2\otimes v_1\la{cap}\,.
\ea
Let us first check the skein relation \eqref{q-dim2} by applying the cap and cup operators in succession:
\bee
1~\rightarrow~-A v_1\otimes v_2+A^{-1}v_2\otimes v_1~\rightarrow~-A(A)+A^{-1}(-A^{-1})=-(A^2+A^{-2})\,.
\eee
We now verify the l.h.s. of \eqref{Id} (the r.h.s. follows in a similar way); the graphical notation has to be understood as: we start with $v_{1,2}\cdot 1$, apply the cap to 1 and end up with  $v_{1,2}\otimes \alpha\otimes \beta$ with $\alpha, \beta\in V$ and finally apply the cup to the $v_{1,2}\otimes  \alpha$ part to end up with a vector in V, namely
\baa
&&\begin{array}{c}\includegraphics[width=5.cm]{id.eps}\end{array}:V~\rightarrow ~ V\otimes V\otimes V~\rightarrow ~V\\
&&v_1~\rightarrow ~-A v_1\otimes v_1\otimes v_2+A^{-1}v_1\otimes v_2\otimes v_1~\rightarrow ~
-A\cdot 0\cdot v_2+A^{-1}A v_1=v_1\\
&&v_2~\rightarrow ~-A v_2\otimes v_1\otimes v_2+A^{-1}v_2\otimes v_2\otimes v_1~\rightarrow ~
-A (-A^{-1}) v_2+A^{-1}\cdot 0\cdot v_1=v_2\,.
\eaa

Therefore, the actions \eqref{cup}, \eqref{cap} fulfill the desired properties. Let us now compute the action of the crossing operators \eqref{cross2a} on the basis of the tensor product vector space $V\otimes V$. This is where the $\su_q(2)$ algebra structure becomes clear. We concentrate on the crossing $\begin{array}{c}\includegraphics[width=.5cm]{x-quantum-2.eps}\end{array}$ and use the r.h.s. of the second skein relation in \eqref{cross2a}, together with the actions \eqref{cup}, \eqref{cap} to see what the crossing does to the basis vectors. We have
\begin{align}\la{braiding}
&\begin{array}{c}\includegraphics[width=1.cm]{x-quantum-2.eps}\end{array}=A\begin{array}{c}
\includegraphics[width=1.cm]{qL1.eps}\end{array}+A^{-1}  \begin{array}{c}
\includegraphics[width=1.cm]{qL2.eps}\end{array}:V\otimes V~\rightarrow ~ V\otimes V\n\\
&v_1\otimes v_1~\rightarrow ~Av_1\otimes v_1\n\\
&v_1\otimes v_2~\rightarrow ~Av_1\otimes v_2+A^{-1}(-A^2 v_1\otimes v_2+v_2\otimes v_1)=A^{-1}v_2\otimes v_1\n\\
&v_2\otimes v_1~\rightarrow ~Av_2\otimes v_1+A^{-1}(v_1\otimes v_2-A^{-2}v_2\otimes v_1)=
A^{-1}v_1\otimes v_2+A^{-1}(A^2-A^{-2})v_2\otimes v_1\n\\
&v_2\otimes v_2~\rightarrow ~Av_2\otimes v_2\,.
\end{align}
We now want to show that the action above of the crossing operator corresponds exactly to the action of the $\su_q(2)$ $R$-matrix in the spin-1/2 (2-dim) representation on $\C^2\otimes\C^2$. Let us first recall a few facts about $\su_q(2)$ (see for instance~\cite{MajidFound} or~\cite{Chari}).

Let $q=e^h$ and let $\su_q(2)$ be the algebra generated by $X_+,X_-,e^{h H}$ with relations
\be\la{q-algebra}
X_+X_--X_-X_+=\frac{e^{2 h H}-e^{-2h H}}{q-q^{-1}}\,,~~~~~e^{h H}X_+=qX_+e^{h H}\,,~~~~~e^{h H}X_-=q^{-1}X_-e^{h H}\,,
\ee
i.e. the q-deformation of the $\su(2)$ algebra in Cartan-Weyl basis.
We then obtain a bi-algebra given by the co-products
\be\la{co-products}
\Delta X_+= X_+\otimes e^{h H}+e^{-h H}\otimes X_+\,,~~~~~\Delta X_-= X_-\otimes e^{h H}+e^{-h H}\otimes X_-\,,~~~~~\Delta e^{h H}=e^{h H}\otimes e^{h H}\,,
\ee
antipodes
\be\la{antipodes}
S\left(H\right)=-H, \qquad
S\left(X_{\pm}\right)=-e^{\pm h}X_{\pm}\,,
\ee
and co-units
\be\la{co-units}
e(X_+)=e(X_-)=0\,,~~~~~e(e^{h H})=1\,.
\ee
This gives a quasi-triangular bi-algebra with $R\in \su_q(2)\otimes \su_q(2)$ given by
\be\la{R}
R=\sum_{n=0}^\infty\frac{q^{\frac{n}{2}(n+1)}(1-q^{-2})^n}{[n]_q !}e^{2h(H\otimes H)}X_+^n\otimes X_-^n\,,
\ee
where
\be
[n]_q=\frac{q^n-q^{-n}}{q-q^{-1}}\,.
\ee
The 2-dimensional representation $\rho$ of $\su_q(2)$, in which $X_+,X_-,H$ act as linear transformations on $\C^2$ is given by
\be
\rho(X_+)=\begin{pmatrix}
  0 & 1 \\
  0 & 0
 \end{pmatrix}\,,~~~~~
 \rho(X_-)=\begin{pmatrix}
  0 & 0 \\
  1 & 0
 \end{pmatrix}\,,~~~~~
\rho(H)=\begin{pmatrix}
  \frac{1}{2} & 0 \\
  0 & -\frac{1}{2}
 \end{pmatrix}\,,
 \ee
 and
 \be
\rho(e^{hH})=\begin{pmatrix}
  e^{\frac{h}{2}} & 0 \\
  0 & e^{-\frac{h}{2}}
 \end{pmatrix}\,.
\ee
Notice that, in this fundamental representation, the algebra \eqref{q-algebra} reduces to the standard (non-deformed) $\su(2)$ algebra; this, again, signals that the algebra itself is not very indicative of the symmetry deformation.

The action of the $R$-matrix \eqref{R} on $V\otimes V$ provides a representation of the non-commutative holonomies braiding \eqref{cross2a}. More precisely, the crossing of two non-commutative holonomies can be represented as the action of the $R$-matrix on the tensor product vector space of two copies of $V$, followed by the switch of the two vector spaces.
Diagrammatically we have
\be
\begin{array}{c}\includegraphics[width=2.5cm]{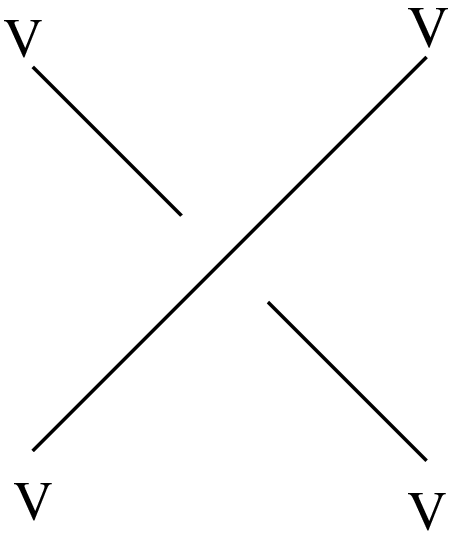}\end{array}=\begin{array}{c}\includegraphics[width=2.6cm]{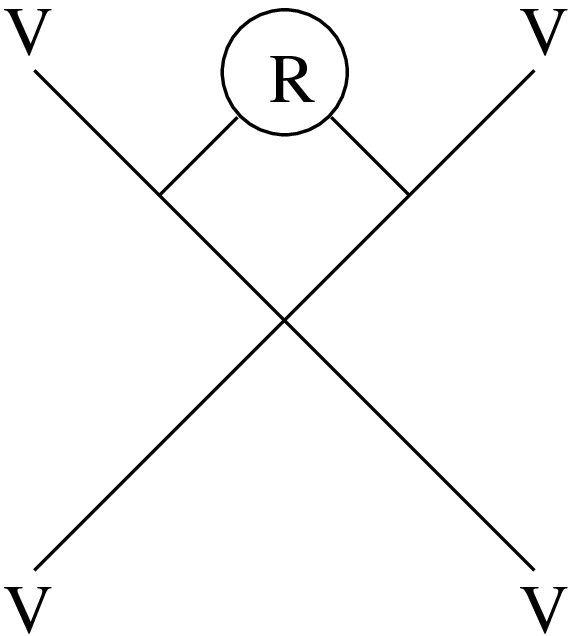}\end{array}\,.
\ee
Let us show this explicitly by computing the action of $\rho(R)$ on a basis of $V\otimes V$ expressed in terms of the $v_1, v_2$ basis of $V=\C^2$ and compare it with \eqref{braiding}. In general, if we express $R=\sum_{i,j} R^{ij}G_i\otimes G_j$, with $G_i$ any basis for $\su_q(2)$, its action on a given vector $v\otimes v'\in V\otimes V'$ is given by
\be
\rho(R)(v\otimes v')=\sum_{ij}R^{ij}S(\rho(G_i) v\otimes \rho(G_j) v')=\sum_{ij}R^{ij} \rho(G_j) v'\otimes \rho(G_i) v~~\in V'\otimes V\,,
\ee
where $S$ is the `switch' operator. Therefore, the action of the 2-dimensional representation of the $R$-matrix \eqref{R} on $\C^2\otimes \C^2$ in terms of a basis $v_1, v_2\in \C^2$ with $\rho(X_+)v_1=\rho(X_-)v_2=0, \rho(E)v_2=v_1, \rho(X_-)v_1=v_2, \rho(H)v_1=1/2v_1, \rho(H)v_2=-1/2v_2$ is given by (notice that only the terms up to $n=1$ in \eqref{R} contribute in the 2-dim representation)
\begin{align}\la{Raction}
&\rho(R)(v_1\otimes v_1)= S(\rho(e^{2h(H\otimes H)})v_1\otimes v_1)= e^{\frac{h}{2}}v_1\otimes v_1\,,\n\\
&\rho(R)(v_1\otimes v_2)= S(\rho(e^{2h(H\otimes H)})v_1\otimes v_2)=e^{-\frac{h}{2}}v_2\otimes v_1\,,\n\\
&\rho(R)(v_2\otimes v_1)= S(\rho(e^{2h(H\otimes H)})v_2\otimes v_1+(e^h-e^{-h})\rho(e^{2h(H\otimes H)}X_+\otimes X_-)v_2\otimes v_1)\n\\
&=e^{-\frac{h}{2}}v_1\otimes v_2+(e^h-e^{-h})e^{-\frac{h}{2}}v_1\otimes v_2\,,\n\\
&\rho(R)(v_2\otimes v_2)= S(\rho(e^{2h(H\otimes H)})v_2\otimes v_2)= e^{\frac{h}{2}}v_2\otimes v_2\,.
\end{align}
We see that, for
\be\la{h}
A=e^{\frac{h}{2}}\,,\quad h=i \hbar \frac{\sqrt{\Lambda}}{2\kappa}\,,
\ee
the actions \eqref{braiding} and \eqref{Raction} coincide.
Therefore, the crossing \eqref{cross2a} of two non-commutative holonomies, in terms of which the curvature constraint is written, can be represented in terms of the generators $X_+, X_-, e^{h H}$ of the quasitriangular bialgebra $\su_q(2)$ through its $R$-matrix.


Let us summarize: we found that the $R$-matrix associated to the crossing of two operators corresponding to the generators $\hat{H}^+[N_p]$ is the $R$-matrix of the Hopf algebra $\su_q(2)$ with $q=\exp(i\hbar \sqrt{\Lambda}/2\kappa)$.
By the same argument, the $R$-matrix associated to the crossing of two operators corresponding to the generators $\hat{H}^-[N_p]$ is the $R$-matrix of the Hopf algebra $\su_{q'}(2)$ with $q'=\exp(-i\hbar \sqrt{\Lambda}/2\kappa)$, see~(\ref{cross2a}),(\ref{cross2b}).  
The fact that the deformation parameters of these two sectors are inverse to each other is remarkable and, as we will see below, crucial for the existence of the non-singular $\Lambda \rightarrow 0$ limit (see~\cite{Celeghini:1990xx}).
As illustrated above, this relation between the deformation parameters of the two sectors follows from recasting the constraint operators as functionals of the noncommutative connections $A^+$ and $A^-$, i.e. it 
 is enforced directly by the structure of the isometry group of the classical space-time and the application of  quantization techniques of the LQG formalism.

Recalling (see Eq.~(\ref{LR})) that the left chiral $\su(2)$ generators ${\cal L}_i$ correspond to $H^+[\delta_i]/\sqrt{\Lambda}$, while the right chiral $\su(2)$ generators ${\cal R}_i$ correspond to $H^-[\delta_i]/\sqrt{\Lambda}$, in the following we will denote by $H^L,X^L_+,X^L_-$ the $\su_{q}^L(2)$ generators associated to the quantization of $H^+$, and by $H^R,X^R_+,X^R_-$ the $\su_{q}^R(2)$ generators associated to the quantization of $H^-$. The (complex) deformation parameters of the left and right copies are related by
\begin{equation}\label{qLqR}
q_L = q^{-1}_R = q =\exp(i\hbar \sqrt{\Lambda}/2\kappa) \ .
\end{equation}

\section{From $\su_q(2)\oplus\su_{q^-1}(2)$ to $\so_q(4)$}
\label{fromSO2toSO4}

We have explicitly shown how the introduction of a regulator, in the form of a discerte structure, required by the LQG quantization scheme, leads to a quantum deformation of the local $\su(2)\oplus \su(2)$ symmetry generated by the classical constraint algebra \eqref{cl-algebra}. At the quantum level, the local isometry becomes\footnote{Notice that the deformation parameter is complex, with $|q| = 1$, so that $q^{-1}=q^*$, and we could also write the algebra as $\su_q(2)\oplus \su_{q^*}(2)$
.} $\su_q(2)\oplus \su_{q^{-1}}(2)$, where we have taken into account  Eq.~(\ref{qLqR}).
At the same time we have shown in Sec.~\ref{symmetries} how the diffeomorphism constraints generating the isometries of $S^3$, closing the Lie algebra $\so(4)$, are related to the curvature constraints $H^\pm[\delta_{(i)}]$ generating the local $\su(2)\oplus \su(2)$ symmetry.

Starting now from $\su_q(2)\oplus \su_{q^{-1}}(2)$, we can reverse the procedure in order to find the deformed algebra of the symmetry generators of $\so_q(4)$.
In this way we will obtain the (Hopf) algebra describing the relativistic symmetries of the deformed de Sitter space-time with Euclidean signature. From now on we will set $\hbar=1$.

Recall first that the maps~(\ref{KillingSO4SU2}) define the relations between the $\mathfrak{so}(4)$ generators~(\ref{SO4}) and the chiral  $\mathfrak{su}(2)\oplus\mathfrak{su}(2)$ ones~(\ref{SU2SU2}) to be
\begin{equation}\label{SO4SU2}
\begin{gathered}E=-\frac{\sqrt{\Lambda}}{2}\left({\cal L}_3+{\cal R}_3\right),\qquad P_{a}=-\frac{\sqrt{\Lambda}}{2}\epsilon_{ab}\left({\cal L}_b-{\cal R}_b\right),\\
M=\frac{1}{2}\left({\cal L}_3-{\cal R}_3\right),\qquad
N_{a}=-\frac{1}{2}\left({\cal L}_a+{\cal R}_a\right),
\end{gathered}
\end{equation}
where remember that ${\cal L}_i = H^+[\delta_i]/\sqrt{\Lambda}$ and ${\cal R}_i = H^-[\delta_i]/\sqrt{\Lambda}$.
In order to recover $\so_q(4)$ from $\su_q(2)\oplus \su_{q^{-1}}(2)$, we first need the relation between the left and right chiral generators ${\cal L}_i$, ${\cal R}_i$, and the Cartan-Weyl generators $H^L,X_\pm^L$ and $H^R,X_\pm^R$ characterizing the two copies of $\su_{q}^L(2)$ and $\su_{q}^R(2)$, each one described by Eqs. (\ref{q-algebra}), (\ref{co-products}), (\ref{antipodes}).
This map is given, for each of the two copies, by
\begin{equation}\label{SL2SU2}
\begin{gathered}
H^{L}=\frac{i}{2}{\cal L}_{3},\qquad X_{\pm}^{L}=\frac{i}{2}\left({\cal L}_{1}\pm i{\cal L}_{2}\right),\\
H^{R}=\frac{-i}{2}{\cal R}_{3},\qquad X_{\pm}^{R}=\frac{-i}{2}\left({\cal R}_{1}\pm i{\cal R}_{2}\right),
\end{gathered}
\end{equation}
and is such that, if ${\cal L}_i$ and ${\cal R}_i$ close the classical algebra (\ref{SU2SU2}), then $H^L,X_\pm^L$ and $H^R,X_\pm^R$ close two copies of the $\su(2)$ algebra\footnote{Where in that case the coproducts are primitive ($\Delta X = X\otimes 1 + 1 \otimes X$), as customary for a standard Lie algebra.} in the Cartan-Weyl basis
\begin{equation}\label{HX}
\left[H^{i},H^{j}\right]=0,\qquad\left[H^{i},X_{\pm}^{j}\right]=\pm\delta_{ij}X_{\pm}^{j},\qquad\left[X_{+}^{i},X_{-}^{j}\right]=2\delta_{ij}H^{j},  \qquad i,j=L,R \ .
\end{equation}
Composing (\ref{SO4SU2}) with (\ref{SL2SU2}) we finally obtain the relations between the $\so(4)$ generators (\ref{SO4}) and the $\su(2)$ generators in Cartan-Weyl basis (\ref{HX}):
\begin{equation}\label{mapSL2-SO4}
\begin{gathered}E=i\sqrt{\Lambda}\left(H^{L}-H^{R}\right),\qquad M=-i\left(H^{L}+H^{R}\right),\\
P_{1}=\frac{\sqrt{\Lambda}}{2}\left(X_{+}^{L}-X_{-}^{L}+X_{+}^{R}-X_{-}^{R}\right),\\
P_{2}=-\frac{i\sqrt{\Lambda}}{2}\left(X_{+}^{L}+X_{-}^{L}+X_{+}^{R}+X_{-}^{R}\right),\\
N_{1}=\frac{i}{2}\left(X_{+}^{L}+X_{-}^{L}-X_{+}^{R}-X_{-}^{R}\right),\\
N_{2}=-\frac{1}{2}\left(-X_{+}^{L}+X_{-}^{L}+X_{+}^{R}-X_{-}^{R}\right).
\end{gathered}
\end{equation}

We can now turn to the deformed case.
In the quantum theory, we have seen that the local isometry is codified by two copies $\su_q^L(2)$ and $\su_q^R(2)$, each one being the quantum deformation of \eqref{HX}. For each copy, the algebra is given by Eq.~(\ref{q-algebra}), which we rewrite as
\begin{equation}
\begin{gathered}
\left[H^{i},H^{j}\right]=0, \qquad
\left[H^{i},X_{\pm}^j\right]=\pm\delta_{ij}X_{\pm}^j, \\
\left[X_{+}^i,X_{-}^j\right]=\delta_{ij}\frac{\sinh\left(2h_{i}H^{i}\right)}{\sinh\left(h_{i}\right)}\,,
\label{eq:SUq(2)}
\end{gathered}
\qquad i,j = L,R
\end{equation}
The coalgebra sector is given by Eq.~(\ref{co-products})
\begin{equation}
\begin{gathered}
\Delta H^{i}=H^{i}\otimes1+1\otimes H^{i}, \\
\Delta X_{\pm}^i=X_{\pm}^i\otimes e^{h_{i}H^{i}}+e^{-h_{i}H^{i}}\otimes X_{\pm}^i\,,
\label{eq:SUq(2)co-products}
\end{gathered}
\qquad i,j=L,R
\end{equation}
and the antipodes by Eq.~(\ref{antipodes})
\begin{equation}
S\left(H^{i}\right)=-H^{i}, \qquad
S\left(X_{\pm}^i\right)=-e^{\pm h_{i}}X_{\pm}^i\,,
\qquad i,j=L,R\,.
\label{eq:SU(2)antipodes}
\end{equation}

Notice that, in our case (see Eq.~\eqref{qLqR}), 
\be\la{z}
h_L=-h_R=h=i z\,,\quad z= \sqrt{\Lambda}/2\kappa \,,
\ee
(i.e. $q_L=q_R^{-1}=e^h$), with $h$ given in \eqref{h}. The opposite sign of $h_L$ and $h_R$ will play a crucial role in the discussion, presented in the next section, of the $\Lambda\rightarrow 0$ contraction limit of the deformed algebra.

We will use now the maps (\ref{mapSL2-SO4}) together with relations (\ref{eq:SUq(2)}), (\ref{eq:SUq(2)co-products}) and (\ref{eq:SU(2)antipodes}) to obtain the deformed algebra $\so_q(4)$ as $\so_q(4)\simeq \su_{q_L}(2)\oplus\su_{q_R}(2) = \su_q(2)\oplus\su_{q^{-1}}(2)$.
Using (\ref{mapSL2-SO4}) with (\ref{eq:SUq(2)}), the algebraic part of $\mathfrak{so}_{q}\left(4\right)$ is then
\begin{equation}\label{algebraSOq4}
\begin{gathered}\left[E,P_{a}\right]=\Lambda N_{a},\qquad\left[N_{a},E\right]=P_{a},\\
\left[P_{1},P_{2}\right]=\Lambda\frac{\sinh\left(zM\right)}{\sin\left(z\right)}\cosh\left(zE/\sqrt{\Lambda}\right),\\
\left[N_{a},P_{b}\right]=-\delta_{ab}\sqrt{\Lambda}\frac{\sinh\left(zE/\sqrt{\Lambda}\right)}{\sin\left(z\right)}\cosh\left(zM\right),\\
\left[N_{1},N_{2}\right]=\frac{\sinh\left(zM\right)}{\sin\left(z\right)}\cosh\left(zE/\sqrt{\Lambda}\right),\\
\left[M,N_{a}\right]=\epsilon_{a}^{\ b}N_{b},\qquad\left[M,P_{a}\right]=\epsilon_{a}^{\ b}P_{b},\qquad\left[M,E\right]=0.
\end{gathered}
\end{equation}
Notice that the algebraic part does not depend on the relative sign
of $h_{L}$ and $h_{R}$ (we would have obtained the same commutators
choosing $h_{L}=h_{R}=h$). This is not true for the coalgebraic part
of $\mathfrak{so}_{q}\left(4\right)$. Using (\ref{mapSL2-SO4}),
(\ref{eq:SUq(2)co-products}) and (\ref{eq:SU(2)antipodes}), we find
\begin{equation}\label{co-productsSOq4}
\begin{gathered}\Delta E=E\otimes1+1\otimes E\ ,\qquad\Delta M=M\otimes1+1\otimes M\ ,\\
\begin{split}\Delta P_{a}= & P_{a}\otimes e^{\frac{1}{2}zE/\sqrt{\Lambda}}\cosh\left(\frac{1}{2}zM\right)+e^{-\frac{1}{2}zE/\sqrt{\Lambda}}\cosh\left(\frac{1}{2}zM\right)\otimes P_{a}\\
 & +\epsilon_{ab}\,\left(\sqrt{\Lambda}N_{b}\otimes e^{\frac{1}{2}zE/\sqrt{\Lambda}}\sinh\left(\frac{1}{2}zM\right)-\sqrt{\Lambda}e^{-\frac{1}{2}zE/\sqrt{\Lambda}}\sinh\left(\frac{1}{2}zM\right)\otimes N_{b}\right)\,,
\end{split}
\\
\begin{split}\Delta N_{a}= & N_{a}\otimes e^{\frac{1}{2}zE/\sqrt{\Lambda}}\cosh\left(\frac{1}{2}zM\right)+e^{-\frac{1}{2}zE/\sqrt{\Lambda}}\cosh\left(\frac{1}{2}zM\right)\otimes N_{a}\\
 & -\epsilon_{ab}\,\left(\frac{1}{\sqrt{\Lambda}}P_{b}\otimes e^{\frac{1}{2}zE/\sqrt{\Lambda}}\sinh\left(\frac{1}{2}zM\right)-\frac{1}{\sqrt{\Lambda}}e^{-\frac{1}{2}zE/\sqrt{\Lambda}}\sinh\left(\frac{1}{2}zM\right)\otimes P_{b}\right)\,,
\end{split}
\end{gathered}
\end{equation}
\begin{equation}\label{antipodeSOq4}
\begin{gathered}S\left(M\right)=-M\ ,\qquad S\left(E\right)=-E\ ,\\
S\left(P_{a}\right)=-\cos\left(z\right)P_{a}-\sqrt{\Lambda}\sin\left(z\right)N_{a}\ ,\\
S\left(N_{a}\right)=-\cos\left(z\right)N_{a}+\frac{1}{\sqrt{\Lambda}}\sin\left(z\right)P_{a}\,.
\end{gathered}
\end{equation}
If we had the same sign for the two deformation parameters,
$h_{L}=h_{R}=h$, we would obtain the same co-products but with
the role of $E/\sqrt{\Lambda}$ and $M$ inverted. This will turn out to
be crucial for the convergence of the co-product in the contraction
limit.

It is worth pointing out that taking the limit $z\rightarrow 0$, one recovers the $\mathfrak{so}(4)$ Lie algebra (\ref{SO4}) (with primitive coproducts and antipodes), as it should be. It is clear then that our deformed algebra describes a genuine deformation of the $\so(4)$ algebra described in Sec.~(\ref{symmetries}).

Let's now pause for a moment and discuss the reality conditions of the deformed algebra $\so_q(4)$ that we have found. These determine the $*$-operation defined on the deformed algebra (an antilinear anti-involution corresponding to a complex conjugation), and the unitarity of the $R$-matrix.
Notice first that the $\so_q(4)$ generators $E$, $P_a$, $N_a$, $M$ are anti-Hermitian operators, as  is clear from their commutation relations and the fact that conjugation extends as an anti-algebra map.
It follows from relations~(\ref{mapSL2-SO4}) that the generators $H^i$ and $X_\pm^i$ ($i=L,R$) of $\su_q(2)$ ($\su_{q^-1}(2)$) satisfy the reality conditions:
\begin{equation}\label{reality}
\left(H^i\right)^* = H^i,\qquad \left(X_+^i\right)^* = X_-^i,\qquad \left(X_-^i\right)^* = X_+^i,\qquad i=L,R \ .
\end{equation}
This justifies the name $\su_q(2)$ ($\su_{q^-1}(2)$) for the corresponding algebras.
The fact that $|q|=1$ implies that we we have to deal with the case discussed in~\cite{BorowiecEuclidean,BorowiecLorentzian,tolstoyFlip}, where the $*$-operation is lifted to the tensor product $\so_q(4) \otimes \so_q(4)$ by the ``flip''
\begin{equation}\label{flip}
(a \otimes b)^* = b^*\otimes a^* \ ,
\end{equation}
so that
$\so_q(4) \simeq \su_q(2) \oplus \su_{q^-1}(2)$ is endowed with a (flipped) $*$-algebra homomorphism\footnote{Here $\tau(a\otimes b) = b\otimes a$ is the flip operation, while $(a\otimes b)^{*\otimes*}=a^*\otimes b^*$.}
\begin{equation}
(\Delta h)^* := \tau\circ(\Delta h)^{*\otimes*} = \Delta(h^*) \qquad h \in \so_q(4) \ .
\end{equation}
Under such ``flipped'' $*$-operation the $R$ matrix of $\su_q(2) \oplus \su_{q^-1}(2)$ is unitary in the sense that
\begin{equation}
R^*=\tau(R^{*\otimes*}) = R^{-1},
\end{equation}
as it can be shown by considering its related classical $R$-matrix obtained by taking the linear order of $R$ in $h$ ($q=\exp(h)$ and $R=R_{\su_q(2)}R_{\su_{q^{-1}}(2)} \simeq 1 + r$, with $R_{\su_q(2)}$ given by Eq.~(\ref{R})):
\begin{equation}
r =  2h\left(H^{L}\otimes H^{L}-H^{R}\otimes H^{R}+X_{+}^{L}\otimes X_{-}^{L}-X_{+}^{R}\otimes X_{-}^{R}\right) .
\end{equation}
Then, under the flipped $*$-operation, using Eqs.~(\ref{reality}) and (\ref{flip}), $r^* = -r$ for $h$ purely imaginary, as it is in our case, and $R$ is unitary.

We close this section by observing that one can obtain the Lorentzian version $\so_q(3,1)$ of our deformed algebra by a Wick rotation\footnote{This has to be intended as a formal transformation that does not affect reality of the generators and constants.} $E\rightarrow -iE$, $N_a\rightarrow -iN_a$, $z\rightarrow iz$.
It is worth noting that in this case the deformation parameter $q$ becomes real, and the $\so_q(3,1)$ splits in a direct sum $\su_q(2) \oplus \su_{q^-1}(2)$ of two (mutually commuting) copies of $\su_q(2)$, with a (flipped) $*$-structure that interchanges the two factors
\begin{equation}
\begin{gathered}
(H^L)^* = H^R,\qquad (X_+^L)^* = X_-^R,\qquad (X_-^L)^* = X_+^R,\\
(H^R)^* = H^L,\qquad (X_+^R)^* = X_-^L, \qquad (X_-^R)^* = X_+^L \ ,
\end{gathered}
\end{equation}
as in the complexification procedure discussed for instance in~\cite{MajidFound} Sec.~7.3.

\section{From $\so_q(4)$ to $\kappa$-Poincar\'e}

In this section, we derive the $\kappa$-Poincar\'e algebra, co-products and antipodes from the Inonu-Wigner contraction \cite{Inonu} of \eqref{algebraSOq4}, \eqref{co-productsSOq4} and \eqref{antipodeSOq4}.
The contraction is performed by substituting $z=\sqrt{\Lambda}/2\kappa$
and then taking the limit $\sqrt{\Lambda}, z\rightarrow0$ while keeping $\kappa = \sqrt{\Lambda}/2z$ finite. We get
\begin{equation}\label{algebra-cont}
\begin{gathered}\left[E,P_{a}\right]=\left[P_{1},P_{2}\right]=0,\qquad\left[N_{a},E\right]=P_{i},\\
\left[N_{a},P_{b}\right]=-\delta_{ab}\kappa\sinh\left(E/\kappa\right),\qquad\left[N_{a},N_{b}\right]=M\cosh\left(E/\kappa\right),\\
\left[M,N_{a}\right]=\epsilon_{ab}\, N_{b},\qquad\left[M,P_{a}\right]=\epsilon_{ab}\, P_{b},\qquad\left[M,E\right]=0\,,
\end{gathered}
\end{equation}
\begin{equation}\label{co-product-cont}
\begin{gathered}\Delta E=E\otimes1+1\otimes E\ ,\qquad\Delta M=M\otimes1+1\otimes M\ ,\\
\begin{split}\Delta P_{a}= & P_{a}\otimes e^{\frac{1}{2}E/\kappa}+e^{-\frac{1}{2}E/\kappa}\otimes P_{a}\,,\end{split}
\\
\Delta N_{a}=N_{a}\otimes e^{\frac{1}{2}E/\kappa}+e^{-\frac{1}{2}E/\kappa}\otimes N_{a}
-\frac{1}{2\kappa}\epsilon_{ab}\,\left(P_{b}\otimes e^{\frac{1}{2}E/\kappa}M-e^{-\frac{1}{2}E/\kappa}M\otimes P_{2}\right)\,,
\end{gathered}
\end{equation}
\begin{equation}\label{antipode-cont}
\begin{gathered}S\left(M\right)=-M,\qquad S\left(E\right)=-E,\\
S\left(P_{a}\right)=-P_{a},\qquad S\left(N_{a}\right)=-N_{a}+\frac{1}{\kappa}P_{a}\,.
\end{gathered}
\end{equation}
Equations \eqref{algebra-cont}, \eqref{co-product-cont}, \eqref{antipode-cont} define the (2+1)D $\kappa$-Poincar\'e algebra in standard basis and
Euclidean signature.

It is at this point that the fact that $h_L$ and $h_R$ have opposite sign becomes crucial. As said before, had we had $h_{L}=h_{R}=h$,
the role of $E/\sqrt{\Lambda}$ and $M$ in the co-products (\ref{co-productsSOq4})
would have been inverted.
One can easily see that performing the contraction $\sqrt{\Lambda}, z\rightarrow0$ with $\kappa= \sqrt{\Lambda}/2z$ finite in (\ref{co-productsSOq4}), the co-products of the boosts $N_{a}$
would have  diverged.
The role of the opposite sign of $h$ for the two copies in the convergence of the contraction was first noticed in~\cite{Celeghini:1990xx}.
As pointed out at the end of Sec.~\ref{sec:$R$-matrix}, this is indeed the case for the two $R$-matrices associated to the generators $\hat H^+, \hat H^-$ corresponding to the left and right $\su_q(2)$ copies. Therefore, the relations (\ref{z}) that we obtained consistently with the LQG quantization of the constraint algebra of the theory, are also the ones that guarantee the convergence of the contraction.

It is a straightforward but tedious calculation to show that if in~(\ref{algebraSOq4}),(\ref{co-productsSOq4}) and~(\ref{antipodeSOq4}) we change the generators as~(see\cite{Majid:1994cy} and \cite{Amelino})
\begin{equation}
\begin{gathered}E=\tilde{E},\qquad M=\tilde{M},\qquad P_{a}=e^{z\tilde{E}/\left(2\sqrt{\Lambda}\right)}\tilde{P}_{a},\\
N_{a}=e^{z\tilde{E}/\left(2\sqrt{\Lambda}\right)}\left(\tilde{N}_{a}-\frac{z}{2\sqrt{\Lambda}}\epsilon_{ab}\,\tilde{M}\tilde{P}_{b}\right),
\end{gathered}
\end{equation}
then the contracted algebra becomes the (2+1)D $\kappa$-Poincar\'e algebra in the bicrossproduct
basis and Euclidean signature (after removing the tildes):
\begin{equation}
\begin{gathered}\left[E,P_{a}\right]=\left[P_{1},P_{2}\right]=0,\qquad\left[N_{a},E\right]=P_{a},\\
\left[N_{a},P_{b}\right]=-\delta_{ab}\left(\frac{\kappa}{2}\left(1-e^{-2E/\kappa}\right)-\frac{1}{2\kappa}\vec{P}^{2}\right)-\frac{1}{\kappa}P_{a}P_{b},\\
\left[N_{1},N_{2}\right]=M,\qquad\left[M,N_{a}\right]=\epsilon_{ab}\, N_{b},\qquad\left[M,P_{a}\right]=\epsilon_{ab}\, P_{b},\qquad\left[M,E\right]=0.
\end{gathered}
\end{equation}
\begin{equation}
\begin{gathered}\Delta E=E\otimes1+1\otimes E\ ,\qquad\Delta M=M\otimes1+1\otimes M\ ,\\
\begin{split}\Delta P_{a}= & P_{a}\otimes1+e^{-E/\kappa}\otimes P_{a}\,,\end{split}
\\
\Delta N_{a}=N_{a}\otimes1+e^{-E/\kappa}\otimes N_{a}
-\frac{1}{\kappa}\epsilon_{ab}\, P_{b}\otimes M\,,
\end{gathered}
\end{equation}
\begin{equation}
\begin{gathered}S\left(M\right)=-M,\qquad S\left(E\right)=-E,\\
S\left(P_{a}\right)=-e^{E/\kappa}P_{a},\qquad S\left(N_{a}\right)=-e^{E/\kappa}N_{a}+\frac{1}{\kappa}\epsilon_{ab}\, e^{E/\kappa}P_{b}M\,.
\end{gathered}
\end{equation}
Again, if $E\rightarrow -iE$, $N_a\rightarrow -iN_a$, $\kappa \rightarrow -i\kappa$, we recover the $\kappa$-Poincar\'e Hopf algebra in Lorentzian signature.

\section{Coupling to massive point particles}\la{Particles}

We have shown that the local isometry $\so(4)\simeq\su(2)\oplus\su(2)$ of classical 3D gravity with a positive cosmological constant is deformed at the quantum level, where the space-time local symmetry becomes the quantum group $\so_q(4)$. 

We can now imagine  coupling massive  point particles to the theory. We are not going to present a technical analysis of the problem, but only discuss the possible implications of our results presented above in the presence of particles. This section does not contain any new results, but its motivation is to highlight possible implications and, at the same time,  to emphasize further the physical relevance of the deformed quantum space-time symmetries we derived in the previous sections.

It is well known \cite{Deser, Witten, deSousa, Matschull} that massive particles represent topological degrees of freedom, introducing conical singularities at their location with deficit angle $\theta= 4\pi G m$ proportional to the mass $m$ of the particle. However, it can be shown that massive particles have no gravitational interaction in three dimensions, i.e. they do not affect the geometry of space-time. This aspect represents a crucial departure from the four dimensional case and it allows for a different approach to the quantization of the system. In order to clarify this point, let us first recall some basic elements of the inclusion of particles in 3D gravity (see, for instance, \cite{deSousa,Matschull,Bais:2002ye,Meusburger:2003ta,Meusburger:2005mg,AGN,NP2} and references therein).

The main relevant feature to our discussion represents the fundamental role played by the isometry algebra of gravity. More precisely, one can describe the coupling of a point  particle to classical gravity in the Chern-Simons approach evolving in an homogenous 3D space-time using an algebraic formalism. In the case of our interest (Riemannian gravity with $\Lambda>0$), the homogenous space is given by the sphere $S^3$, which can be seen as the coset $G/H$, with $G=SO(4)$ and $H=SU(2)$. The degrees of freedom of the particle are given by a group element $g\in G$, which, by means of the map $G\rightarrow G/H \times H$, can be decomposed into $g=qp\rightarrow (q,p)$. Then, $q\in G/H$ can be associated to the position of the particle in space-time, while $p\in H$ is associated to its momentum. In this way, the degrees of freedom of a relativistic particle at rest are encoded in two half-integer parameters $(m,s)$ labeling the unitary irreducible representations  of the isometry group $G$ and entering the values of the two Casimir operators of $\so(4)$, written in terms of the generators $P_i, J_i$. The half-integer $m$ is interpreted as the mass of the particle and $s$ as its spin.

The equation of motion derived from the algebraic action can be shown to imply geodesic equations for the position variable $q$ on $G/H$. Moreover, the invariance of the action under (global) left multiplication of the isometry group on itself have the physical interpretation of the theory being invariant under a change of inertial frame. This leads to the appearance of the momentum $p_i$ and the total angular momentum $j_i$ conserved Noether charges for the particle, satisfying the Poisson algebra
\be
\{j_i, j_j\}=\epsilon_{ij}\,^k j_k\,,\quad \{p_i, p_j\}=-\epsilon_{ij}\,^k j_k\,,\quad \{j_i, p_j\}=-\epsilon_{ij}\,^k p_k\,.
\ee
In particular, the quadratic Casimir relations
\be
p_ip^i-j_i j^i=m^2-s^2\,,\quad p_ij^i=ms
\ee
defining the particle mass and spin show how the dynamics of a free particle with mass $m$ and spin $s$ on the homogenous space $G/H\sim S^3$ is correctly encoded in the algebraic action formalism.

The configuration space of the point particle is given by the coadjoint orbits of the $\so(4)$ Lie algebra
\be
\mathcal C_{m,s}\equiv \{g=h(mJ_0+sP_0)h^{-1}\,|\, h\in G\}\,,
\ee
with the Cartan subalgebra generators $J_0, P_0$ defining the rest frame of the particle. When coupling the particle action to gravity, it can be shown that the interaction term can be gauged away and the particle evolves as a free particle. As pointed out above, this indicates the insensitivity of massive point particles to the gravitational field in three dimensions.

In the absence of particles the non-commutative Chern-Simons connection, which can be written as in \eqref{conn}, is flat everywhere. If particles are present, the conical singularities they induce on the space-like surface $\Sigma$ are completely encoded in the conjugacy class of the group elements corresponding to the holonomies of the Chern-Simons connection on small loops going around the punctures. The gauge group $C^\infty(\Sigma, G)$ on the space of regular $G$-connections on $\Sigma$ has an adjoint action also on the coadjoint orbits $C_{m,s}$ of the particles. It is this extended simultaneous action on the whole phase-space of gravity plus punctures which provides an effective coupling between flat connections and particles degrees of freedom.
This effective coupling is a crucial aspect. In fact, it implies that, at the quantum level, the mass and the spin of  particles are given by unitary irreducible representations of the isometry group of quantum geometry.

In the context of canonical combinatorial quantization of Chern-Simons theory, this has been shown in \cite{Alekseev}, allowing to construct the kinematical Hilbert space of the theory in terms of a tensor product of unitary irreducible representations of $SL_q(2)$.
In the context of LQG quantization of Riemannian 3D gravity with $\Lambda=0$, a similar construction has been performed in \cite{Noui}. However, in the case of a vanishing cosmological constant, the presence of particles is crucial for the Drinfeld double symmetry to emerge. This can be understood from the fact that the local conical singularities induced by the particles play a role analogous to the presence of a local constant curvature given by a non-vanishing cosmological constant.

In fact, in our analysis, when $\Lambda>0$ we have shown that, even when particles are not present, loop quantization induces a deformation of the isometry group of classical 3D gravity encoded in the replacement of the $SU(2)$ recoupling theory with the $SL_q(2)$ one and leading to the appearance of the Drinfeld double $DSL_q(2)$ isometry. Since in 3D matter and gravity are not really interacting, one could imagine to couple classical massive point particles to the quantum background geometry defined by the LQG physical Hilbert space defined in \cite{DP3}. The arguments above would then motivate the expectation that
 n-particle states transform as representations of $\so_q(4)$ under
rotations and translations, leading to deformed transformations laws with respect to the classical
$\so(4)$ ones.
In other words, the local  isometry of geometry gets deformed at the quantum level and, through the induced gauge action of gravity on coupled point particles, this would lead to a classification of particles in terms of unitary irreducible representations of such deformed isometry group.

A low energy  regime of the theory could then be investigated
by removing (integrating out) the quantum gravity effects, which effectively amounts to taking the $\Lambda\rightarrow 0$ limit (while keeping $\ell_{\va P}$ finite). This could generally lead to a deformation of gravity classical isometry group in the Minkowski regime.  Indeed, as we saw in the previous section, such contraction leads to the appearance of a $\kappa$-Poincar\'e deformed symmetry. This suggests that, at the Planck scale, the effective theory for the matter sector would correspond to a non-commutative quantum field theory symmetric under the $\kappa$-Poincar\'e group. This scenario was realized in \cite{Freidel}, by applying the covariant formalism of LQG to Riemannian 3D gravity with $\Lambda=0$ coupled to a scalar matter field. Our analysis provides further evidence of this picture from the canonical approach.

\section{Conclusions}

We have studied the LQG quantization of the off-shell constraint algebra of 2+1 gravity with a positive cosmological constant. By rewriting the constraints in terms of holonomies of the non-commutative connection \eqref{conn} we have unraveled the quantum group structure arising from the discrete, extended structure at the core of the LQG kinematical Hilbert space construction. We have shown how, in the low energy regime where the gravitational field is constant and back-reaction can be ignored, the contraction performed by sending the cosmological constant to zero leads to a deformation of the Euclidean flat space-time symmetries, encoded in the $\kappa$-Poincar\'e algebra.

The idea of a Poincar\'e algebra modification following from a linear (lapse and shift functions) limit of deformed hypersurface-deformation algebra of general relativity through LQG quantization techniques was originally proposed in \cite{Bojowald}. However, in  \cite{Bojowald} it was argued that in 4D the $\kappa$-Poincar\'e algebra cannot be recovered from the LQG corrections in general since these do not affect the spatial diffeomorphisms sector of the algebra. Nevertheless, we have shown in Sect.\ 3 and 4 that this is not the case in 3D, namely diffeomorphisms can be expressed in terms of non-commutative holonomies and these are exactly the generators that lead to a deformed constraint algebra symmetry. Whether a  $\kappa$-Poincar\'e symmetry can indeed be derived also in the 4D case or not using techniques analogous to those introduced in \cite{DP2, DP3}   is hard to say, since the extension to four dimensions is highly non-trivial from a technical point of view (see \cite{AmelinoLQG} for a recent alternative attempt to circumvent the obstruction found in \cite{Bojowald}).

Despite these technical difficulties, we expect that the results obtained in this paper might have important consequences for symmetries of quantum spacetime in physical 4 space-time dimensions. Namely (see \cite{Freidel:2003sp} for details of the argument), a planar system in 4 dimensions is described by 3D gravity, but at the same time is a configuration of 4D gravity. It follows that the symmetries of flat 4D quantum spacetime should contain somehow the symmetries of the 3D one. But as we shown the latter is deformed, and therefore the former must be deformed too. Of course this argument should be confirmed by detailed calculations within 4D quantum gravity. The work in this direction is in progress and important insights  might be provided by the results of \cite{Haggard}, where an interesting connection between 4D loop quantum gravity with a cosmological
constant and $SL(2,C)$ Chern-Simons theory in 3D has been discovered.

\section*{Acknowledgements}
We would like to thank  Andrzej Borowiec, Francisco J. Herranz and Jerzy Lukierski for discussions.

For FC, JKG, and GR this work was supported  by funds provided by the National Science Center under the
agreement DEC- 2011/02/A/ST2/00294, and for JKG also by funds
provided by the National Science Center under the agreement
2014/13/B/ST2/04043. DP wishes to acknowledge the Templeton Foundation for the supporting grant number 51876.

\appendix

\section{Diffeomorphisms in 3d gravity}\label{appendix}

In this appendix, we will demonstrate the validity of \eqref{diff}.

The explicit expression of $D[0,\vec{\xi}]={\cal D}[\xi]$, using $e^i_a=-\epsilon_{ab} \,E^b_i$, reads
\begin{equation}
{\cal D}[\xi]=\int_\Sigma d^2x \,\xi^a\left(-2\,F^i_{ab}\, E^b_i + A^i_a\, D_bE^b_i\right)=\int_\Sigma d^2x \,\xi^a\left[(\partial_b A^i_a-\partial_a A^i_b)\, E^b_i + A^i_a\, \partial_bE^b_i\right]\,,
\end{equation}
from which one gets the following transformations for $A^a_i$ and $E^a_i$
\begin{align}
&\delta_{\vec{\xi}} A^i_a= -A^i_b\,\partial_a\xi^b -\xi^b\,\partial_b A^i_a=\mathcal{L}_{\vec{\xi}} A^i_a\label{deltaAsdiff}\\
&\delta_{\vec{\xi}} E_i^a= E_i^b\,\partial_b\xi^a -\xi^b\,\partial_b E_i^a-E^a_i\,\partial_b\xi^b=\mathcal{L}_{\vec{\xi}} E_i^a\,,\label{deltaEsdiff}
\end{align}
and they coincide with the Lie derivatives of $A^i_a$ and $E^a_i$ along $\vec{\xi}$ (note that $E^a_i$ is a density and that's the reason why the term $-E^a_i\,\partial_b\xi^b$ is present in \eqref{deltaEsdiff}).

For $\underline{\xi}=(\xi,\vec{0})=\xi\,\underline{t}$, one has
\begin{equation}
D[\xi\,\underline{t}]=\int_\Sigma \,\xi\left[e_t^i\left(F^i_{ab}\,\epsilon^{ab}+\frac{\Lambda}{2}\,\epsilon^{ijk}\,E^a_j\,E^b_k\,\epsilon_{ab}\right)  + A^i_t\, D_aE^a_i\right]\,,
\end{equation}
from which one gets
\begin{equation}
\delta_{\xi\,\underline{t}} A^i_a= -D_a (\xi\,A^i_t) = -A^i_t\,\partial_a\xi-\xi\,D_a A^i_t-\Lambda\,\xi\,\epsilon^{ijk}\,e^j_t\,E^b_k \,\epsilon_{ab}=-A^i_t\,\partial_a\xi-\xi\,\partial_t A^i_a=\mathcal{L}_{\xi\,\underline{t}} A^i_a\label{deltaAtrep}\\
\end{equation}
where we used the equation of motion $2F_{ta}=\partial_t A^i_a-D_a A^i_t=-\Lambda\, \epsilon_{ijk}\, e^j_t\, e^k_a$.
Similarly for $E^a_i$ one has
\begin{align}
\delta_{\xi\,\underline{t}} E^a_i&= \epsilon^{ba}\, D_b (\xi\,e^i_t) +\xi\, \epsilon_{ijk} A^j_t \, E^a_k =- \epsilon^{ab}\,e^i_t\,\partial_b \xi- \xi\left(\epsilon^{ab}\,D_b e^i_t-\epsilon_{ijk} A^j_t \, E^a_k\right)=\nonumber\\
&=- \epsilon^{ab}\,e^i_t\,\partial_b \xi- \xi\,\epsilon^{ab}\left(D_b e^i_t-\epsilon_{ijk} A^j_t \, e^k_b\right)=- \epsilon^{ab}\,e^i_t\,\partial_b \xi- \xi\,\epsilon^{ab}\,\partial_t e^i_b=\epsilon^{ab} \mathcal{L}_{\xi\underline{t}} e^i_b\label{deltaEtrep}
\end{align}
where we used the equation of motion $D_a e^i_t = D_t e^i_a$ and the relation $E^a_i=\epsilon^{ab}\, e_b^i$.

The equations \eqref{deltaAsdiff}, \eqref{deltaEsdiff}, \eqref{deltaAtrep} and \eqref{deltaEtrep} provide an outline to show that \eqref{diff} generates diffeomorphisms $x^\mu\rightarrow x^\mu+\xi^\mu$ in phase-space.

Let us now consider a 2+1 splitting of the metric tensor: given generic coordinates $y^\mu$ a family of spatial hypersurfaces is defined by
\begin{equation}
y^\mu=y^\mu(t,x^a)\,,
\end{equation}
$x^a$ being coordinates on each hypersurface, while $t$ is a parameter labeling each hypersurface. The deformation vector is defined as
\begin{equation}
\frac{d\underline{y}}{dt}=N\,\underline{n}+N^a\,\underline{b}_a\,,\label{def}
\end{equation}
$\underline{n}$ and $\underline{b}_a$ being the normal and tangential vectors to spatial hypersurfaces, respectively. $N$ and $N^a$ are the lapse function and the shift vector. The metric in coordinates $(t,x^a)$ reads
\begin{equation}
g_{\mu\nu}=\left(\begin{array}{cc} sN^2+N^aN_a & N_a \\ N_a & h_{ab} \end{array}\right)\,,
\end{equation}
$h_{ab}$ being the spatial metric and $N_a=h_{ab}\, N^b$, while $s=+1,-1$ for Euclidean and Lorentzian space-times, respectively. The inverse metric is given by
\begin{equation}
g_{\mu\nu}=\left(\begin{array}{cc} s\frac{1}{N^2} & -s\frac{N^a}{N^2} \\ -s\frac{N^a}{N^2} & h^{ab}+s\frac{N^aN^b}{N^2} \end{array}\right)\,.
\end{equation}
From \eqref{def} it follows that the normal vector $\underline{n}$ has the following components
\begin{equation}
\underline{n}=\left(\frac{1}{N}, -\frac{N^a}{N}\right)\,,
\end{equation}
such that one can write
\begin{equation}
n^\mu=s\,N\, g^{t\mu}\,. \label{ncomp}
\end{equation}

The constraint ${\cal D}[\vec{f}]$ generates spatial diffeomorphisms $\xi^\mu=(0,\vec{f})$, thus from \eqref{diff}
\begin{equation}
{\cal D}[\vec{f}]=D[0,\vec{f}]=C_\Lambda[f^a\,e^i_a]+G[f^a\,A_a^i]\,.\label{sm}
\end{equation}
The constraint ${\cal H}[g]$ generates those diffeomorphisms orthogonal to spatial hypersurfaces, thus along $\underline{n}$. Hence,
\begin{equation}
{\cal H}[g]=D[g\,\underline{n}]=C_\Lambda[g\,n^\mu\,e^i_\mu]+G[g\,n^\mu\,A_\mu^i]\,,
\end{equation}
which using \eqref{ncomp} can be rewritten as
\begin{equation}
{\cal H}[g]=C_\Lambda[s\,g\,N\,e_i^t]+G[s\,g\,N\, g^{t\mu}\,A_\mu^i]\,.\label{sh}
\end{equation}
It can be shown that \eqref{sm} and \eqref{sh} generate the standard constraints algebra
\begin{align}
&\big[{\cal D}[f_1],\, {\cal D}[f_2]\big]={\cal D}\big[[f_1,\, f_2]\big]\\
&\big[{\cal D}[f],\, {\cal H}[g]\big]={\cal H}[f^a\partial_ag]\\
&\big[{\cal H}[g_1],\, {\cal H}[g_2]\big]={\cal D}[f(g_1,g_2)]
\end{align}
where
\begin{equation}
[f_1,\, f_2]=f_1^a\,\partial_a\vec{f}_2-f_2^a\,\partial_a\vec{f}_1\qquad
f^a(g_1,g_2)=h^{ab}(g_1\,\partial_bg_2-g_2\,\partial_bg_1)\,,
\end{equation}
$h^{ab}$ being the inverse spatial metric. In particular, the following identities, relating metric components with momenta $E_i^a$, are to be used:
\begin{align}
&e^t_i=\frac{1}{2\sqrt{sg}}\epsilon_{ijk}\, E^a_j E^b_k\,\epsilon_{ab}\,,\quad g^{tt}=e^t_i e^t_i\,,\quad N=\frac{1}{\sqrt{sg^{tt}}}\\
&e^a_i=\frac{1}{\sqrt{sg}}\epsilon_{ijk}\, e^j_t E^a_k\,,\quad g^{ta}=e^t_i e^a_i\\
&\sqrt{sg}=\frac{1}{2}\epsilon_{ijk}\, e^i_t E^a_j E^b_k\,\epsilon_{ab}\,\quad h_{ab}=e^i_a e^i_b\,.
\end{align}

\end{document}